\documentclass[5p, sort&compress]{elsarticle}

\makeatletter
\def\@author#1{\g@addto@macro\elsauthors{\normalsize%
    \def\baselinestretch{1}%
    \upshape\authorsep#1\unskip\textsuperscript{%
      \ifx\@fnmark\@empty\else\unskip\sep\@fnmark\let\sep=,\fi
      \ifx\@corref\@empty\else\unskip\sep\@corref\let\sep=,\fi
      }%
    \def\authorsep{\unskip,\space}%
    \global\let\@fnmark\@empty
    \global\let\@corref\@empty  %% Added
    \global\let\sep\@empty}%
    \@eadauthor={#1}
}

\def\ps@pprintTitle{%
 \let\@oddhead\@empty
 \let\@evenhead\@empty
 \def\@oddfoot{}%
 \let\@evenfoot\@oddfoot}
\makeatother

\usepackage{placeins}
\usepackage{hyperref}
\usepackage{lineno}
\modulolinenumbers[5]
\usepackage{subcaption}
\usepackage{float}
\usepackage{graphicx}
\usepackage{amssymb}
\usepackage{amsmath}
\usepackage{color}
\newcommand{\cg}{\textnormal{\textsl{g}}}

\begin{document}

\begin{frontmatter}
\title{Response of photomultiplier tubes to xenon scintillation light}

\author{B.~L\'opez~Paredes\corref{correspondingauthor}}
\cortext[correspondingauthor]{Corresponding author}
\ead{b.lopez@imperial.ac.uk}
\author{H.~M.~Ara\'ujo}
\author{F.~Froborg}
\author{N.~Marangou}
\author{I.~Olcina}
\author{\\T.~J.~Sumner}
\author{R.~Taylor}
\author{A.~Tom\'as}
\author{A.~Vacheret}
\address{High Energy Physics Group, Imperial College London, Blackett Laboratory, London, SW7 2BW, U.K.}





\begin{abstract}
We present the precision calibration of 35 Hamamatsu R11410-22 photomultiplier tubes (PMTs) with xenon scintillation light centred near $175$~nm. This particular PMT variant was developed specifically for the LUX-ZEPLIN (LZ) dark matter experiment. A room-temperature xenon scintillation cell coupled to a vacuum cryostat was used to study the full-face PMT response at both room and low temperature (\mbox{$\sim\!-100^\circ$C}), in particular to determine the quantum efficiency (QE) and double photoelectron emission (DPE) probability in LZ operating conditions. For our sample with an average QE of (32.4$\pm$2.9)\% at room temperature, we find a \textit{relative} improvement of \mbox{(17.9$\pm$5.2)\%$$} upon cooling (where uncertainty values refer to the sample standard deviation). The mean DPE probability in response to single vacuum ultraviolet (VUV) photons is \mbox{(22.6$\pm$2.0)\%} at low temperature; 
the DPE increase relative to room temperature, measured here for the first time, was \mbox{(12.2$\pm$3.9)\%}. Evidence of a small triple photoelectron emission probability ($\sim\!\!0.6\%$) has also been observed. Useful correlations are established between these parameters and the QE as measured by the manufacturer. The single VUV photon response is also measured for one ETEL D730/9829QB, a PMT with a more standard bialkali photocathode used in the ZEPLIN-III experiment, for which we obtained a cold DPE fraction of \mbox{(9.1$\pm$0.1)\%}. Hence, we confirm that this effect is not restricted to the low-temperature bialkali photocathode technology employed by Hamamatsu. This highlights the importance of considering this phenomenon in the interpretation of data from liquid xenon scintillation and electroluminescence detectors, and from many other optical measurements in this wavelength region.
\end{abstract}


\end{frontmatter}

%\linenumbers

\section{Introduction}
\label{S:Introduction}
Despite recent advances in vacuum ultraviolet (VUV) silicon sensor technology, photomultiplier tubes (PMTs) with quartz windows remain the sensor of choice to detect the VUV scintillation light generated in xenon radiation detectors. Specific models developed for high radiological purity are employed in two-phase (liquid/gas) xenon time projection chambers (LXe-TPCs), which are a leading technology for direct dark matter searches~\cite{Chepel2013}. In these detectors, a significant number of PMTs sense both the prompt scintillation and the electroluminescence signals emitted from the liquid and gaseous phases, respectively. The 3-inch Hamamatsu R11410 PMT is a popular model, offering high quantum efficiency (QE) to xenon VUV light, good low-temperature performance due to a low-resistivity bialkali photocathode (with added bismuth)~\cite{Nakamura2010}, besides extremely low radiological background. Several studies related to its performance in xenon detectors can be found in the literature~\cite{Lung2012,Baudis2013,Lyashenko2014,Baudis2015,Akimov2016,Li2016,Barrow2017}. These are also the sensors adopted for the LUX-ZEPLIN (LZ) dark matter experiment~\cite{LZTDR}, which motivated the study presented here---specifically, the `-22' variant was developed for LZ.

We describe in this article the precision calibration of 35 Hamamatsu R11410-22 PMTs with xenon scintillation light, both in ambient conditions and at low temperature (approximately $-100{\rm^o}$C). We also measured a single unit of another PMT model: the ETEL~D730/9829QB, used in the first science run of ZEPLIN-III~\cite{Akimov2007,Lebedenko2009}; significant characterisation data exist for this model also~\mbox{\cite{Araujo2004,Neves2010}}, and it is interesting to include in this study a different photocathode technology (standard bialkali) from another leading manufacturer. A full acceptance testing programme is being carried out in parallel by LZ collaborators to characterise all (494) R11410 LZ PMTs at room and low temperature using visible light \mbox{(cf.~Section~3.4.1 in Ref.~\cite{LZTDR})}.

There are several technical motivations for this work. Firstly, Hamamatsu measure the VUV response of these phototubes only at room temperature, while the photocathode response for short wavelengths is known to improve upon cooling, and the same is true of the multiplication gain (see, e.g. Ref.~\cite{Araujo2004}, or Ref.~\cite{Lyashenko2014} for this PMT model).

Secondly, the manufacturer's VUV QE calibration is conducted with a small, 5-mm diameter spot size on the photocathode, with illumination provided by a deuterium lamp filtered through a monochromator to a central wavelength of $175$~nm (and other relevant values), with a FWHM of only $4$~nm~\cite{HamamatsuPrivate}; a systematic error of $10\%$ is indicated for this calibration. The scintillation emission spectrum from liquid xenon is centred at $174.8$~nm with a significantly larger FWHM of $10.2$~nm~\cite{FUJII2015293}; the electroluminescence spectrum from gaseous xenon peaks at a somewhat shorter wavelength of $171$~nm with a FWHM of $12$~nm~\cite{Takahashi1983} (it is noted that this is a room temperature measurement). Therefore, a calibration is required that is more representative of the scintillation emitted by liquid and gaseous xenon incident over the whole photocathode.

Finally, and perhaps most importantly, it has been recognised recently that more than one `photoelectron' can be emitted in response to a single VUV photon in some PMT models, including the R11410~\cite{Faham2015}. Although this phenomenon has been known for decades~\cite{Sobieski1976,Johnson1988}, the impact of this double photoelectron emission (DPE) in LXe-TPCs had not been fully appreciated until recently.\footnote{Note that only the primary electron directly released by the photon can be termed the  `photoelectron', with the second one being produced by subsequent impact ionisation within the photocathode layer; we use the term `double photoelectron' loosely, mostly to distinguish these electrons from others generated in the multiplication process.} Hamamatsu report a DC measurement, whereby the QE is determined by dividing the steady photocathode current by the incident VUV photon flux. This $QE^{\scriptscriptstyle\textrm{DC}}$~does not represent the probability that a VUV photon will produce a detectable response ($QE^{\scriptscriptstyle\textrm{P}}$). The latter is the quantity required in experiments which must estimate without bias the stochastic distribution of scintillation photons emitted in particle interaction events. Understanding and quantifying this phenomenon led to a significant improvement in response linearity and in energy resolution in the LUX experiment~\cite{Akerib2016}.

The response of these detectors to very low energy electronic and nuclear recoils cannot be correctly interpreted without this understanding. The single photon response for each PMT  must be determined accurately, including the DPE emission fraction ($\sim$20\%, as measured in Ref.~\cite{Faham2015}) and its temperature dependence, which has not been studied previously.

By testing to percent level a sufficient sample of the LZ phototubes, we aim to establish a good correlation between these response parameters and those specified by the manufacturer. Naturally this information will permit cross-calibration of all (494) units in the LZ TPC and it will be useful to other liquid xenon experiments and beyond. To this end we employed a xenon gas scintillation cell maintained at room temperature illuminating a vacuum cryostat containing seven PMTs, which are cooled to $-97.4^\circ$C, the nominal LZ operating temperature. We measured precisely the absolute QE for xenon scintillation as well as the response to single scintillation photons of the same wavelength, which may involve the generation of one or two photoelectrons and other phenomena.

This paper is organised as follows: we introduce our PMT response model in Section~\ref{sec:theory}; we describe the experimental methodology in Section~\ref{sec:procedures}; we then present the results for the single photon response and the response to the main scintillation pulse in Section~\ref{sec:results}; and we discuss our results in Section~\ref{sec:discussion}. Conclusions are presented in Section~\ref{sec:conclusions}.

\begin{figure}[!hbt]
  \centering  
  \includegraphics[height=9.5cm]{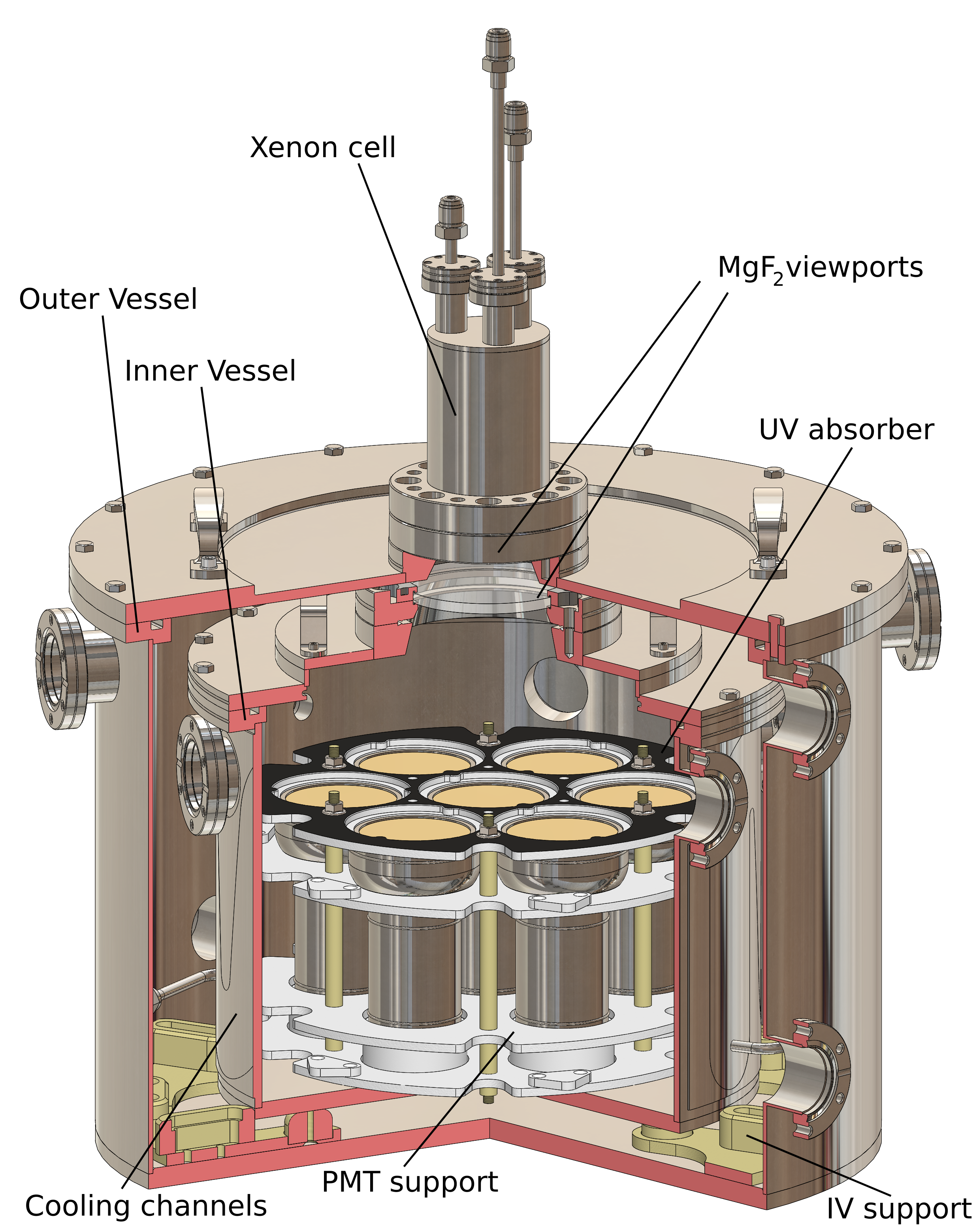}
  \caption{Solid model of the VUV cryostat. The xenon scintillation cell is attached to the outer vessel (OV) lid, illuminating the PMT array through a large MgF$_2$ viewport on the inner vessel (IV) lid. The IV is cooled with liquid nitrogen vapour flowing along meandering channels formed on its external walls.}
\label{fig:hardware}
\end{figure}

%%%%%%%%%%%%%%%%%%%%%%%%%%%%%%%%%%%%%%%%%%%%%%%%%%%%%%%%%%%%%%%%%%%%%%%%%%%

\section{PMT response model}
\label{sec:theory}

When a photoelectron (phe) is produced by a photon incident on the PMT photocathode, it may diffuse to the surface of the sensitive layer and be emitted into the vacuum inside the PMT if it has sufficient energy to overcome the work function of the photosensitive layer. The probability that these three steps are successful is termed the quantum efficiency (QE) of the photocathode~\cite{3stepmodel}. The emitted electron accelerates in the electric field created by the potential $V_\textrm{k-dy1}$ applied between the photocathode and the first dynode. The collection efficiency ($\eta$) represents the probability that the photoelectron reaches the first dynode and successfully multiplies through the dynode chain to produce a measurable electronic signal at the PMT anode. This efficiency $\eta$ depends on the electric field distribution and strength, the geometry and the materials of the electron multiplier, and in particular the design of the first stages of electron multiplication.
Occasionally, for sufficiently short photon wavelengths, the primary photoelectron can lead to the emission of a second electron via impact ionisation in the photocathode, which may result in DPE. In either case, each `photoelectron' can contribute to the signal. The ratio between the charge collected at the anode and the input charge is the gain of the PMT, and it depends on the dynode voltage distribution and electrostatic design.

Various models have been proposed to describe the gain fluctuations experienced by a single photoelectron (SPE) emitted by the photocathode---characterised by the distribution of charge, $q$, measured at the PMT anode for a mean gain $\cg$ and standard deviation $\sigma$, $P(q;\cg,\sigma)$. This charge is often obtained by integrating a digitised waveform and given in pVs. For the Hamamatsu R11410 this distribution is generally found to be well described by a Gaussian function for a sufficiently high gain. In the case of the ETEL D730/9829QB, the SPE distribution is better described by a Polya function~\cite{Neves2010}.

\begin{figure*}[tbh!]
  \centering       
  \includegraphics[width=0.9\textwidth]{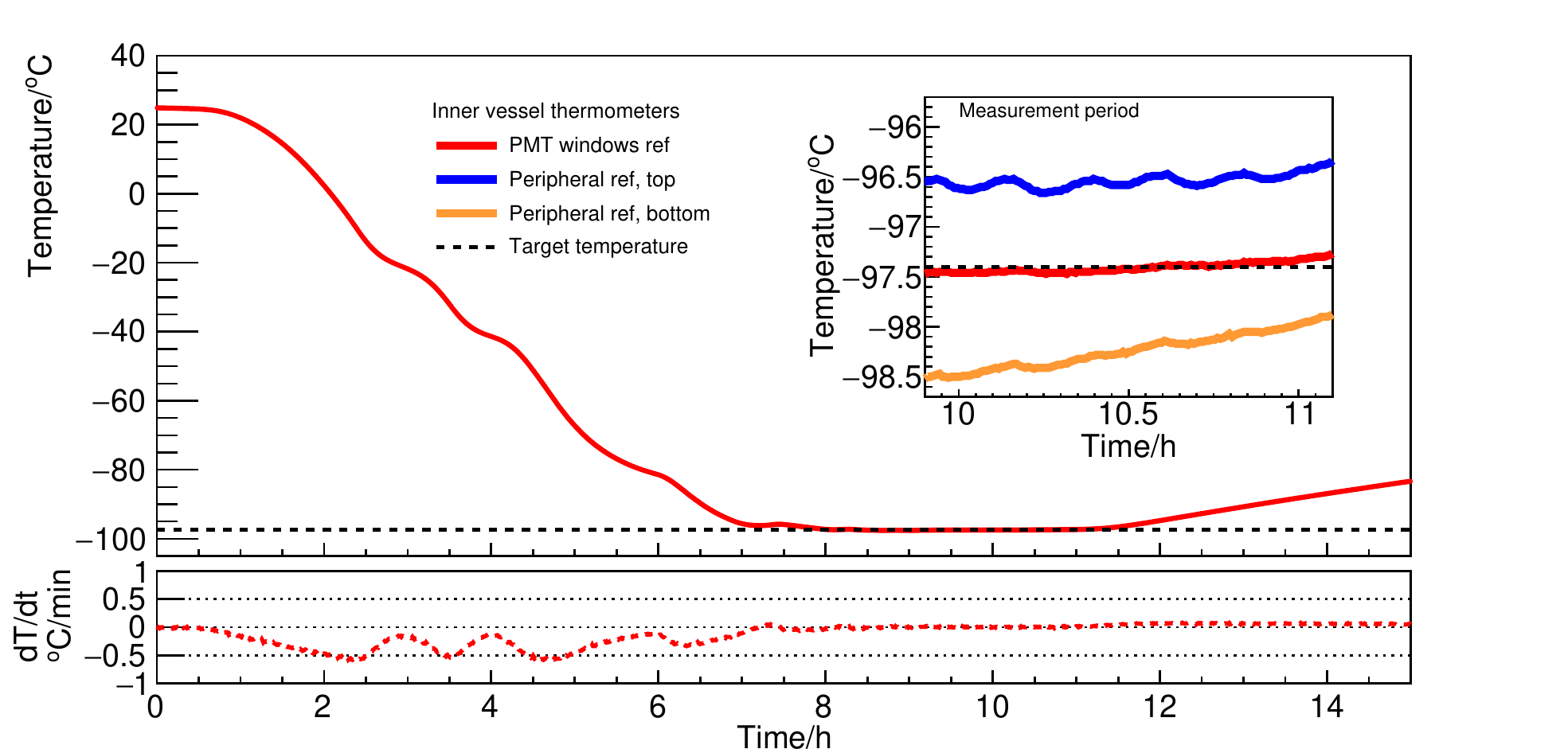}
  \caption{A typical cryostat cooldown curve. The reference temperature for the PMT windows is shown in red and the target temperature of $-97.4^\circ\textrm{C}$ is shown as a dashed black line. The cooling rate is shown in the lower panel as a dashed red line; it remains within the manufacturer specification $|dT/dt|<1^\circ\textrm{C/min}$. The measurement is conducted in the final $\sim1$~h of the stabilisation period, as shown in the inset. Temperatures nearer the cryostat walls are shown in blue and orange.}
  \label{fig:CooldownCurve}
\end{figure*}

In DPE events the gain fluctuations are well modelled assuming that the multiplication of the two electrons is uncorrelated. Quite generally, the single photon response distribution can be described by
\begin{eqnarray}
\label{eq:fitmodel}
  \mathcal{P}(q) &=& (1-f^{\scriptscriptstyle\textrm{DPE}})\,P(q;\cg,\sigma) \notag \\
       &+& f^{\scriptscriptstyle\textrm{DPE}}\,P(q;2\cg,\sqrt{2}\sigma) \notag \\
       &+& c\,P(q;3\cg,\sqrt{3}\sigma)%\,\textrm{,}
\end{eqnarray}
\noindent
for $q>0$, where $f^{\scriptscriptstyle\textrm{DPE}}$ is the DPE \textit{fraction}\footnote{The DPE fraction is thus defined in this study as the number of DPE events divided by the sum of SPE and DPE events. We exclude higher multiple emission from the ratio as it is expected to be very small; our definition of DPE is more readily applicable and has already been used in other experiments~\cite{Faham2015, Akerib2016, PandaX2017}.} and $c$ accounts for the small fraction of events consistent with triple photoelectron emission (TPE). Neglecting the very small $3\!$~-phe contribution, the mean pulse area obtained in response to single VUV photons is
\begin{equation}
\label{eq:gaincalc}
\mu = \langle \mathcal{P}(q;c = 0)\rangle = \int_{0}^{\infty}{q\cdot \mathcal{P}(q;c=0)\,dq}\,\textrm{,}
\end{equation}
which, in the case of a sizeable DPE fraction, will clearly differ from the mean SPE pulse area, $\mu_1$:
\begin{equation}
\mu_1 = \langle \mathcal{P}(q;f = 0, c = 0) \rangle\textrm{.}
\end{equation}

To measure the absolute photocathode QE a source of known photon flux and wavelength is necessary. The DC measurement provided by Hamamatsu ($QE^{\scriptscriptstyle\textrm{DC}}_{\scriptscriptstyle\textrm{H}}$) is subject to several systematic effects when applied to PMT operation in photon-counting mode, including that due to the presence of DPE emission, which generates twice the anode charge for a fraction of photons. This is unlikely to be important for applications where PMTs are used in DC mode, but it is a major effect in photon counting experiments, where the number of detected photons is estimated by explicitly counting the number of pulses in the waveforms. Thus, $QE^{\scriptscriptstyle\textrm{DC}}_{\scriptscriptstyle\textrm{H}}$ can only be used in combination with $\mu_1$ when calculating the number of photons incident on the photocathode. If the DPE fraction for the spectrum of incident light is known, $QE^{\scriptscriptstyle\textrm{DC}}_{\scriptscriptstyle\textrm{H}}$ can be used to derive the true photocathode QE for the photon-counting PMT response:
\begin{equation}
	QE^{\scriptscriptstyle\textrm{P}} = \frac{\mu_1}{\mu}\,QE^{\scriptscriptstyle\textrm{DC}}_{\scriptscriptstyle\textrm{H}} \,\textrm{,}
\end{equation}
which represents the probability that a photon striking the photocathode produces a detectable response.

Given a light pulse inducing a signal with pulse area $A$ at the PMT anode, the absolute number of photons, $N$, incident on the photocathode can then be estimated as
\begin{equation}
\label{eq:nabs}
N = \frac{A}{\eta \; \mu \; QE^{\scriptscriptstyle\textrm{P}}} = \frac{A}{\eta \; \mu_1 \; QE^{\scriptscriptstyle\textrm{DC}}_{\scriptscriptstyle\textrm{H}}} \,\textrm{,}
\end{equation}
where $\eta$ depends on the DPE fraction. We also define the number of \textit{detected photons} and the number of \textit{photoelectrons}, respectively, as
\begin{eqnarray}
\label{eq:nphd}
n_{\scriptscriptstyle\textrm{phd}} &=& A/\mu\textrm{\,,}\\
\label{eq:nphe}
n_{\scriptscriptstyle\textrm{phe}} &=& A/\mu_{1}\textrm{\,.}
\end{eqnarray}

%%%%%%%%%%%%%%%%%%%%%%%%%%%%%%%%%%%%%%%%%%%%%%%%%%%%%%%%%%%%%%%%%%%%%%%%%%%%%%
\section{Measurement procedures}
\label{sec:procedures}

\subsection{Cryostat setup and operation}
\label{sec:cryostat}
A vacuum cryostat was designed and built to cool down the PMTs and maintain their temperature at $-97.4^\circ$C, the LZ operating point---this is illustrated in Figure~\ref{fig:hardware}. The cryostat consists of two stainless steel vessels: an Outer Vessel (OV) and an Inner Vessel (IV). A gaseous xenon scintillation cell, described  below, is mounted onto the OV lid. The IV is supported within the OV by a Delrin\textsuperscript{\textregistered} structure chosen for its low outgassing and small thermal expansion coefficient. A DN100 indium-sealed, MgF$_2$ viewport is attached to the centre of the IV lid and aligned with the xenon cell. MgF$_2$ was also chosen for the xenon cell viewport as it provides a cut-off in transmission well below the wavelengths of interest. Seven upward-facing PMTs rest on a Delrin\textsuperscript{\textregistered} and PEEK frame. A reference PMT is located in the centre, and the measurement is made on the six off-axis units arranged hexagonally, all illuminated by the xenon cell. A set of $70$-mm diameter optical apertures is located above the array. Smaller inserts can be employed to stop down the VUV light, but they were not used in this study; instead, the full $64$~mm photocathode diameter was illuminated. The top of the aperture structure is coated with MetalVelvet\textsuperscript{\tiny TM}, a high-performance UV-absorbing foil.

During operation, the cryostat OV is pumped down to a high vacuum as the xenon VUV radiation is easily absorbed by water vapour~\cite{CHUNG20011572}. We ensure $<\,10^{-5}$~mbar for an effect $\ll\!0.1\%$ on the incident VUV flux. Prior to cooldown, the IV is filled with 1.5~bar(a) N$_2$ gas, which is purified through a SAES FT400-902 getter and serves as the thermal exchange medium. Negligible residual H$_2$O condensation is expected on the PMT windows since the cooling power is delivered through the IV walls, where any condensation will preferentially occur.

Cooling is provided by LN$_2$ vapour delivered from a pressurised dewar. Meandering shapes were made by laser-welding a thin stainless steel sheet on the IV wall, and then hydroformed into channels to allow LN$_2$ vapour to flow. A PID algorithm operates a solenoid valve in the cooling circuit. A typical cooldown uses $<\!10$~litres of LN$_2$.

Eight calibrated Pt100 thermometers were used to monitor the temperature evolution and stability during the measurement. Five thermometers are located along the IV cooling channels and at the bottom IV plate. Another three monitor the inside of the IV: one thermometer is located between the central PMT and an off-axis PMT, and is our reference temperature for the PMT windows; another two are located at the periphery of the array, one at the top and one at the bottom. We ensure that the PMT cooling rate is lower than \mbox{$1^\circ$C/min}, as specified by the manufacturer. A typical cooldown curve is shown in Figure~\ref{fig:CooldownCurve}. The OV and the cell remain strictly at room temperature.

\subsection{Xenon scintillation cell}

In order to study the PMT response in realistic conditions a light source with a spectrum closely resembling that of xenon scintillation and electroluminescence is required. For this purpose a xenon cell was built consisting of a UHV-standard stainless steel chamber capped by a MgF$_2$ viewport. Gas feedthroughs allow for pumping, recirculation and filling of the volume with high purity gas. A SAES PS11-MC1 mini-getter is connected to the cell to allow purification once the system is isolated (in fact, the getter loop was not used for the duration of the PMT testing due to the very long lifetime of the cell).

In order to produce xenon scintillation light a $30$~kBq \mbox{$^{241}$Am} radioactive source is placed in the gas, attached behind a narrow hole in a hemispherical aluminium reflector which also acts as a collimator for the $\alpha$ particles. The reflector shape was modelled so as to optimise the light distribution on the PMTs and the width of the observed spectrum. The collimator reduces the detected $\alpha$-particle count rate to approximately $500$~s$^{-1}$ and narrows the scintillation spectrum for the $4.5$~MeV mean $\alpha$-particle energy.
\begin{figure}[t]
  \centering
  \includegraphics[width=.5\textwidth]{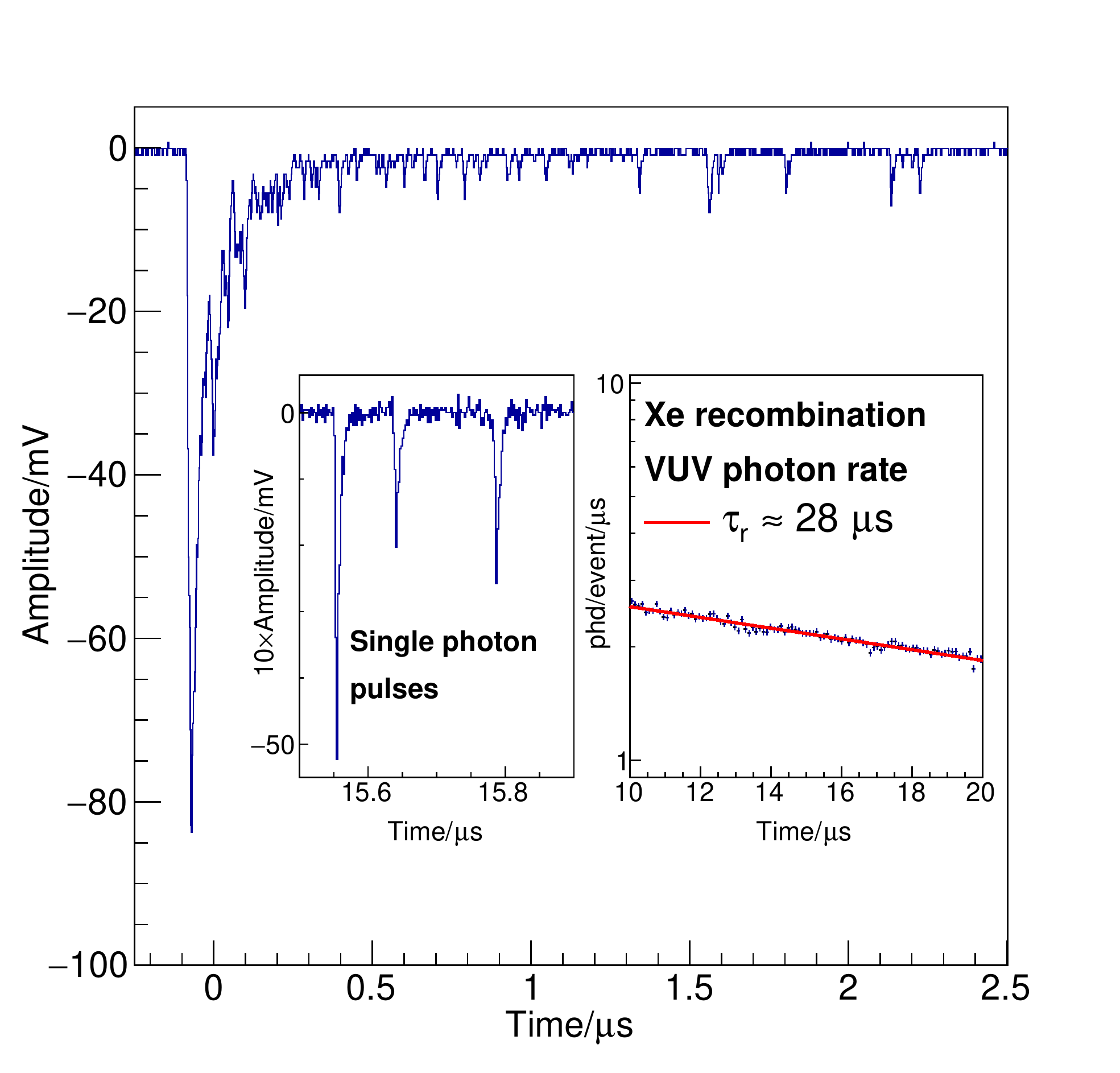}
  \caption{Typical xenon scintillation event due to an $\alpha$-particle interaction. The main (singlet and triplet) components decay within the first few hundred nanoseconds. A possible afterpulsing event appears near \mbox{$t\!\sim\!1.57$~$\mu$s}, likely due to residual \mbox{Ar$^+$}. Electron-ion recombination luminescence with a time constant $\tau_\textrm{r}\!\sim\!28$~$\mu$s is dominant for the remainder of the 20~$\mu$s measurement period. Three single detected photons are shown in the left inset, with one likely DPE emission. On the right inset, the late response measured by an off-axis PMT is shown: $10~\mu$s after the trigger the pulse rate is $<\!3$~phd/$\mu$s/event, and the cell can be used as a source of single VUV photons.}
  \label{fig:Ze3raEvent}
\end{figure}

A typical scintillation event is shown in Figure~\ref{fig:Ze3raEvent}. The decay of the singlet and triplet states of the $\textrm{Xe}_2^\ast$ dimer is observed during the first few hundreds of nanoseconds after the trigger. At later times only the electron-ion recombination luminescence persists. Its time constant is very dependent on the cell geometry, the gas density and purity, and the recombination tail can be very long in the gas phase---as much as hundreds of $\mu$s~\cite{PhysRevA.25.600, Saito2003, mimura2009}; we measure $\tau_r\!\sim\!28$~$\mu$s in our cell. The flux from recombination photons constitutes approximately $26\%$ of the total and it decays exponentially and eventually becomes low enough to be used to our advantage as a source of isolated VUV photons. The level of random coincidences and its effect on the DPE measurement are discussed in Section~\ref{sec:results}.

\begin{figure}[t]
  \centering       
  \includegraphics[width=.5\textwidth]{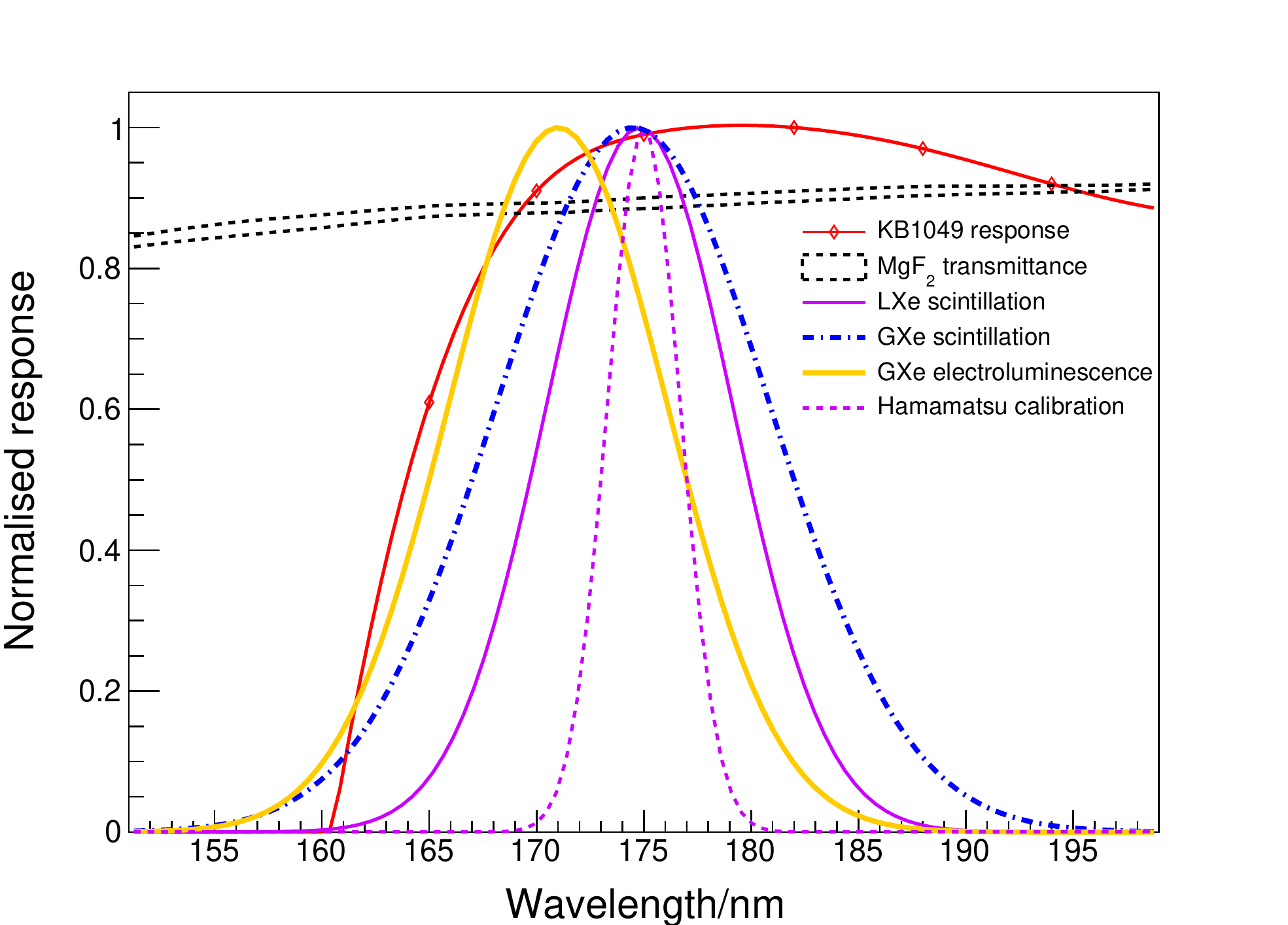}
  \caption{Measured scintillation spectra for liquid (violet)~\cite{FUJII2015293} and gaseous (blue)~\cite{Jortner1965, Saito2002} xenon, along with the (room temperature) gaseous xenon electroluminescence spectrum (yellow)~\cite{Takahashi1983}, MgF$_2$ transmittance (black)~\cite{Jortner1965, mimura2009}, R11410-22 KB1049 PMT relative response (red) and Hamamatsu filtered deuterium lamp spectrum at $175$~nm (violet-dashed).}
  \label{fig:CellVUVSpectrum}
\end{figure}

Published measurements of the liquid and gaseous xenon scintillation spectra are shown in Figure~\ref{fig:CellVUVSpectrum}, together with the MgF$_2$  transmittance. The transmittance is practically flat at xenon scintillation wavelengths, but its energy dependence introduces a small distortion to the spectral shape and consequently a change in the PMT response. The PMT response is estimated using the measurements provided by Hamamatsu at different wavelengths, with a $\textrm{FWHM}=(4\pm2)$~nm. Based on those data, the PMT response to the xenon cell is only $0.2\%$ higher than that for the case of a constant transmittance (i.e. the true scintillation spectral shape), and around $2\%$ lower than the response to liquid xenon scintillation. Given that some of the spectral measurements indicated here are several decades old---and possibly affected by DPE emission systematics themselves---we do not correct for this effect in this study.

Once assembled, the cell was pumped to $4\cdot10^{-9}$~mbar and ultra-high purity gaseous xenon was then circulated for 5~days through a heated SAES MC500 getter, before it was finally filled to 2.5~bar(a) of pure gas. The scintillation yield was monitored for two months with an ETEL D730/9829QB PMT setup. A half-life $>\!500$~days was measured. Prior to starting the LZ PMT testing, the cell was refilled with clean xenon. During the testing the cell half-life was determined to be $>\!1000$~days.

The mean number of VUV scintillation photons reaching the PMTs in the array was estimated using the QE measurements provided by Hamamatsu. The mean of the $\alpha$-particle spectrum corresponds to $\sim\!2000$ photons per decay incident on the central (reference) PMT, and $\sim\!1000$ photons on the off-axis PMTs. More detail on this calibration is given below.

\subsection{Data acquisition}

The ZEPLIN-III data acquisition and analysis frameworks were used to acquire and analyse most data~\cite{ZE3RA}. The DAQ is based on 500~MS/s, 8-bit resolution DC265 Acqiris digitisers arranged in dual-gain configuration --- High Sensitivity (HS) and Low Sensitivity (LS), coupled through 10$\times$ amplifiers (300 MHz Phillips 770). The combined frontend bandwidth is $\sim$150~MHz. The trigger is derived from a pulse-shaped sum of all HS channels. All seven HS and LS channels were recorded per trigger, covering a \mbox{$\sim$21}~$\mu$s waveform duration (including a short pre-trigger region). The early part of this waveform contains the $\alpha$-particle response, while the recombination photons recorded at the later times are used for the calibration of the single VUV photon response as described above.
\begin{figure*}[t]
\begin{subfigure}{0.5\textwidth}
\centering
\includegraphics[width=1.0\textwidth]{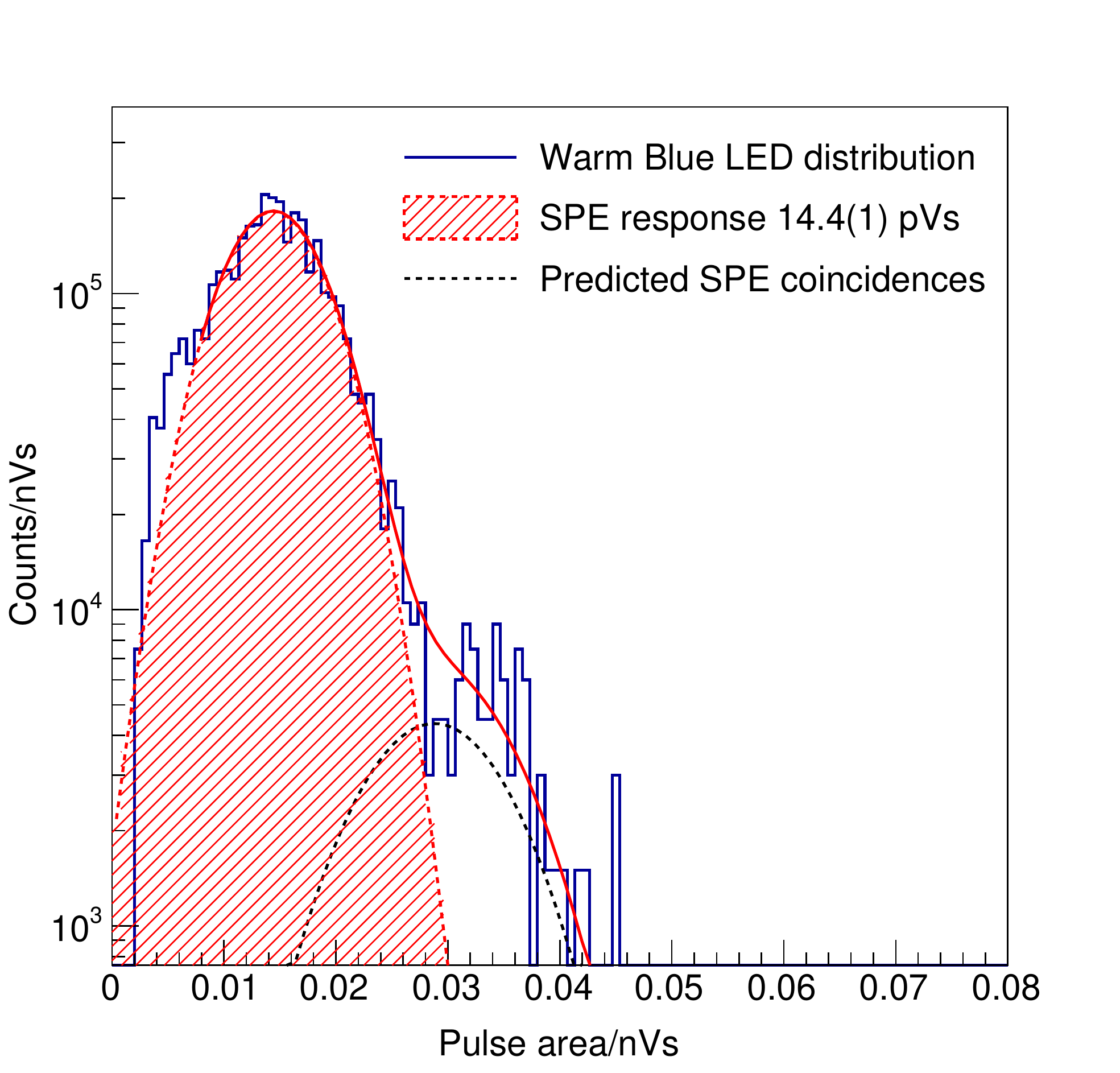}
\end{subfigure}
\begin{subfigure}{0.5\textwidth}
\centering
\includegraphics[width=1.0\textwidth]{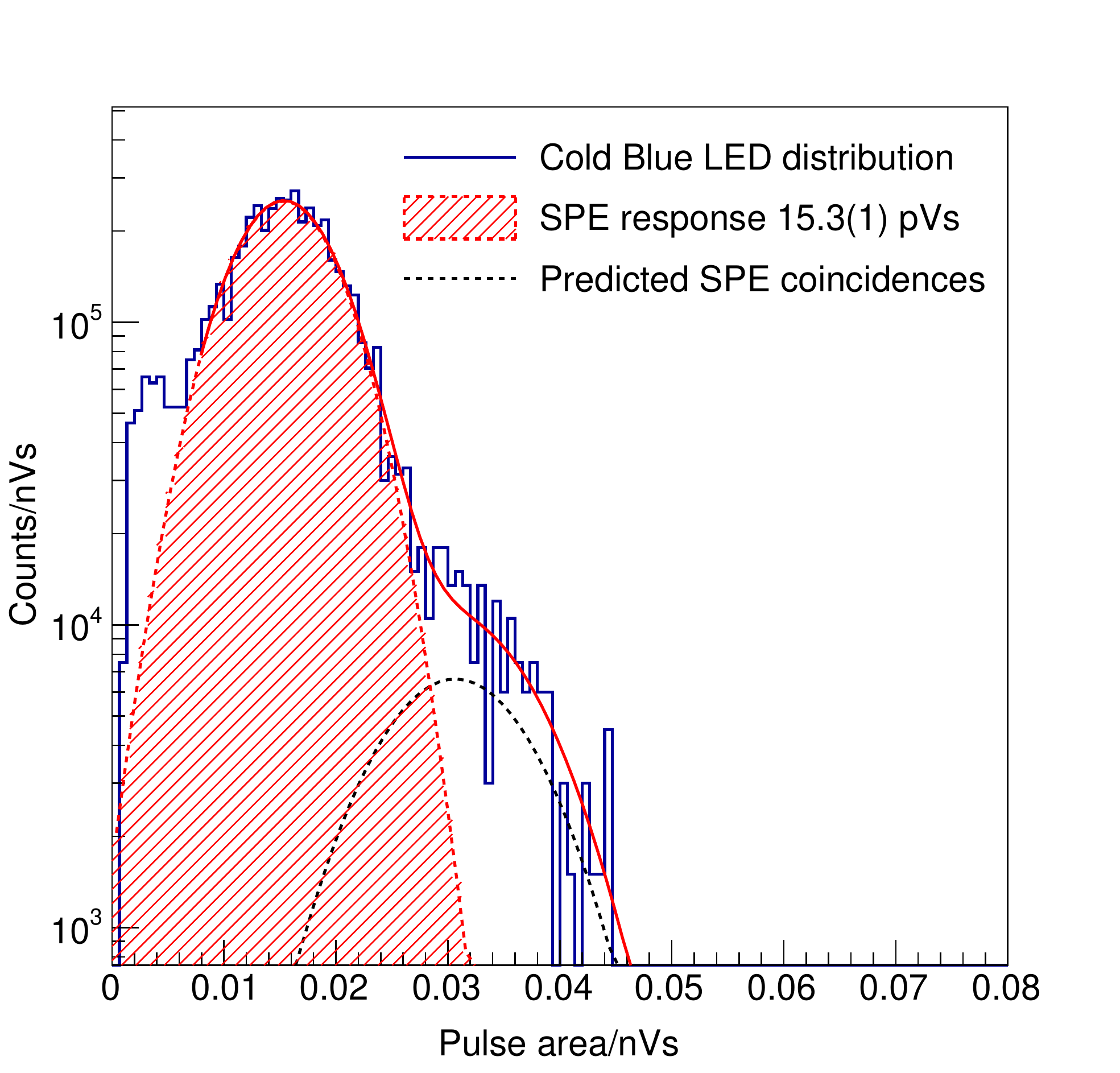}
\end{subfigure}
\caption{Gaussian fit to the pulse area distribution of single blue LED photons for a PMT at room temperature (left) and at $-97.4^\circ\textrm{C}$ (right). The estimated contribution from random coincidences of single photon pulses to the DPE peak is plotted as a dashed black line. This is estimated from the LED and off-trigger SPE rates, and accounts for most of the second peak population.}
\label{fig:BlueLEDGainFitExample}
\end{figure*}

\subsection{Room temperature measurements}

The room temperature measurements were performed in vacuum at a stable temperature of $(23.0\pm0.4)^\circ\textrm{C}$. Prior to each run the cryostat IV and OV were pumped for at least two days, until the pressure reached \mbox{$\lesssim$$10^{-5}$}~mbar. At this pressure, we estimate the effect of any residual water pressure on the VUV measurement to be well below $1\%$, as described in Section~\ref{sec:cryostat}. 

The PMTs were biased to $5\cdot10^6$ gain for at least two hours prior to any measurement using the voltages provided by the manufacturer (in the range 1.30 to 1.53~kV). The voltage divider bases used were those developed for LZ~\cite{LZTDR}, using the voltage distribution recommended by Hamamatsu. At least $6\cdot10^4$ scintillation events were recorded to determine the pulse area with a statistical uncertainty lower than 0.1\%, triggering on the summed signal from the seven PMTs.

An LED pulser system was used to generate blue light pulses ($435$~nm) in order to perform two different measurements. The xenon cell cannot be blinded, but the LED system is externally triggered and these two measurements do not conflict. For the first LED dataset a $\sim\!100$~ns long pulse with $>\!100$~mV amplitude is generated in order to record PMT afterpulsing due to residual gas. This gas, when ionised by photoelectrons in the PMT vacuum, drifts back to the photocathode and generates a delayed signal (see e.g.~\cite{FahamThesis,Barrow2017}). For the second LED measurement the width is $<\!5$~ns (FWHM) and the amplitude is small enough so that each PMT records a signal only every $\sim\!10$ events on average, in order to measure the SPE distribution with $\lesssim\!1\%\,$ 2-phe contamination.

The room temperature measurements (hereafter labelled \textit{warm}) were performed in vacuum to prevent VUV absorption from water outgassing once the IV is filled with nitrogen gas. The VUV measurements were repeated within $20$~min after the IV is pressurised to $1.5$~bar(a) of N$_2$, and no significant differences were found. Thus, no systematic is expected due to performing the cold measurement in N$_2$ gas.

\subsection{Low temperature measurements}

The VUV and LED measurements were repeated once the system had stabilised within \mbox{$(-97.4 \pm0.1)^\circ\textrm{C}$} for at least two hours (typically preceded by another two hours within \mbox{$\sim\!1^\circ\textrm{C}$} of the target temperature). Note that while the PMT envelope and window reach thermal equilibrium with the N$_2$ gas quickly, it takes much longer to thermalise the internal multiplier structures. Therefore, we do not report absolute gain measurements in this study, but the cold measurements are accurate for the photocathode processes (QE and DPE). In some cases, another measurement is conducted after the system has fully recovered to room temperature, and it was found to be consistent with the initial measurement.

%%%%%%%%%%%%%%%%%%%%%%%%%%%%%%%%%%%%%%%%%%%%%%%%%%%%%%%%%%%%%%%%%%%%%%%%%%%%%%%%%
\section{Results}
\label{sec:results}

\subsection{Single-photon response calibration}

\subsubsection{Blue LED response}
\label{sec:blueled}
A dataset with events triggered by the fibre-coupled LED pulser is used to calibrate the PMT response in the absence of DPE, as only a very small DPE fraction is expected at $435$~nm~\cite{Faham2015}. The LED is biased to deliver a mean of $<\!3$~incident photons across the whole PMT array. A time delay histogram is built per PMT to identify the arrival time of the LED photons with respect to the trigger signal. The single photon pulse area distributions thus obtained are fitted and the mean response is extracted, both at room and at low temperature; one example using an R11410 PMT is depicted in Figure~\ref{fig:BlueLEDGainFitExample}.

Two processes account for the small ($\sim$3\%) fraction of 2-phe response observed in these data: i) the probability of random coincidences between blue photons and any other sources of SPE signals (from VUV photons or dark counts, for example) is determined from the off-trigger SPE pulse rate; and ii) the probability for double blue-photon detection is calculated from the frequency of no-response events in the trigger window assuming Poisson statistics.

As the PMT multipliers do not thermalise fully during the cold measurement, some gain drift is expected. These fits are used later on to check the consistency of the VUV results. They also allow us to ensure the validity of the model used to fit the single photon pulse area distribution in conditions where no significant double photoelectron emission is expected.

\subsubsection{VUV photon response}

\begin{figure}[t]
    \centering
    \includegraphics[width=.5\textwidth]{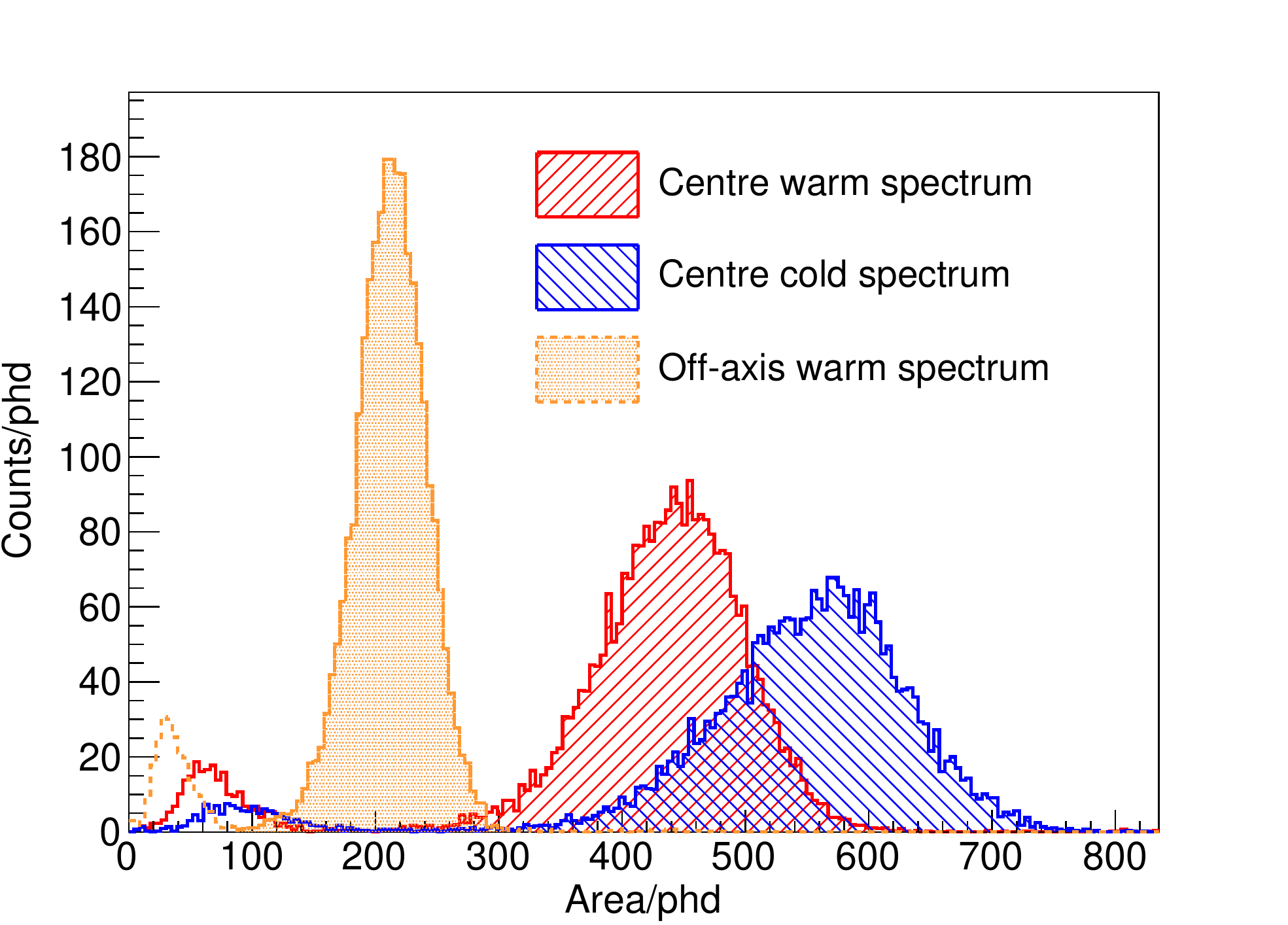}
    \caption{Xenon cell spectra measured by the reference PMT at room temperature (red) and at $-97.4^\circ\textrm{C}$ (blue). The orange-dashed line shows the same interactions measured by an off-axis PMT. The shaded areas indicate the response consistent with an $\alpha$-particle interaction in the xenon; those at lower energies are due to $\gamma$- and X-rays from de-excitation of the $^{237}$Np nucleus.}
    \label{fig:TypicalAlphaSpectra}
\end{figure}

We begin by selecting a population of scintillation events that provide a consistent optical stimulus. The 7-PMT summed pulse areas must be consistent with an $\alpha$-particle interaction in the gaseous xenon and selected events are required to produce a reasonably uniform illumination of the PMT array. This is achieved by accepting events in which the lowest relative pulse area on a PMT is within three standard deviations of the mean summed response of all PMTs. Resulting spectra are shown in Figure~\ref{fig:TypicalAlphaSpectra}.

Then we select single VUV photon candidate pulses. To minimise the probability of random coincidences, small pulses are selected in the range $10$--$20$~$\mu$s after the main scintillation trigger. In a typical xenon scintillation event from the cell, the probability of two-photon coincidences in the outer PMTs becomes $\lesssim$$0.1\%$ after $10$~$\mu$s. Pulses with a mean charge arrival time too large to be consistent with a detected photon ($\tau > 10$~ns) are removed, as well as those consistent with electronic pick-up ($\tau<3$~ns, amplitude~$<1\textrm{~$\mu$V}$). A pulse-level anti-coincidence cut is applied across all PMTs to minimise contamination of the recombination photon population. PMT afterpulsing poses a particular challenge as this leads to small fast signals delayed from their optical stimulus. Well-defined afterpulse delay times allow the rejection of Ar$^+$ and other species using a time window cut derived from the LED measurements---which we find to be in good agreement with those from Ref.~\cite{Barrow2017}. (Most of the residual gas in the PMT vacuum is removed by internal getters, but species such as argon remain present.) The analysis was cross-checked with a 2-$\mu$s inhibit period before every candidate pulse to confirm that any residual afterpulsing at shorter time delays would not impact our results.

\begin{figure*}[t]
\begin{subfigure}{0.5\textwidth}
\centering
\includegraphics[width=1.0\textwidth]{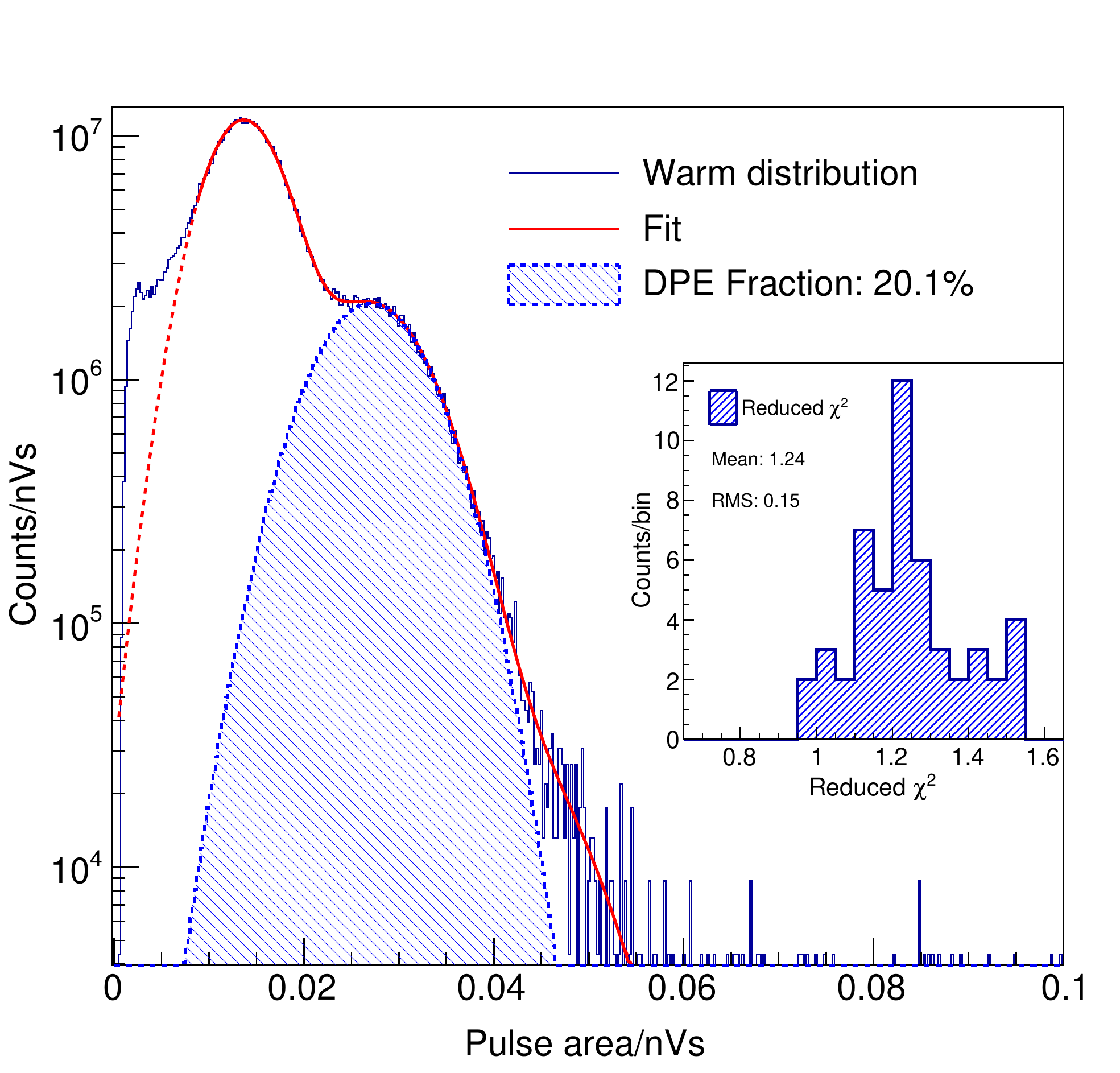}
%\caption{Warm fit}
%\label{fig:WarmFit}
\end{subfigure}
\begin{subfigure}{0.5\textwidth}
\centering
\includegraphics[width=1.0\textwidth]{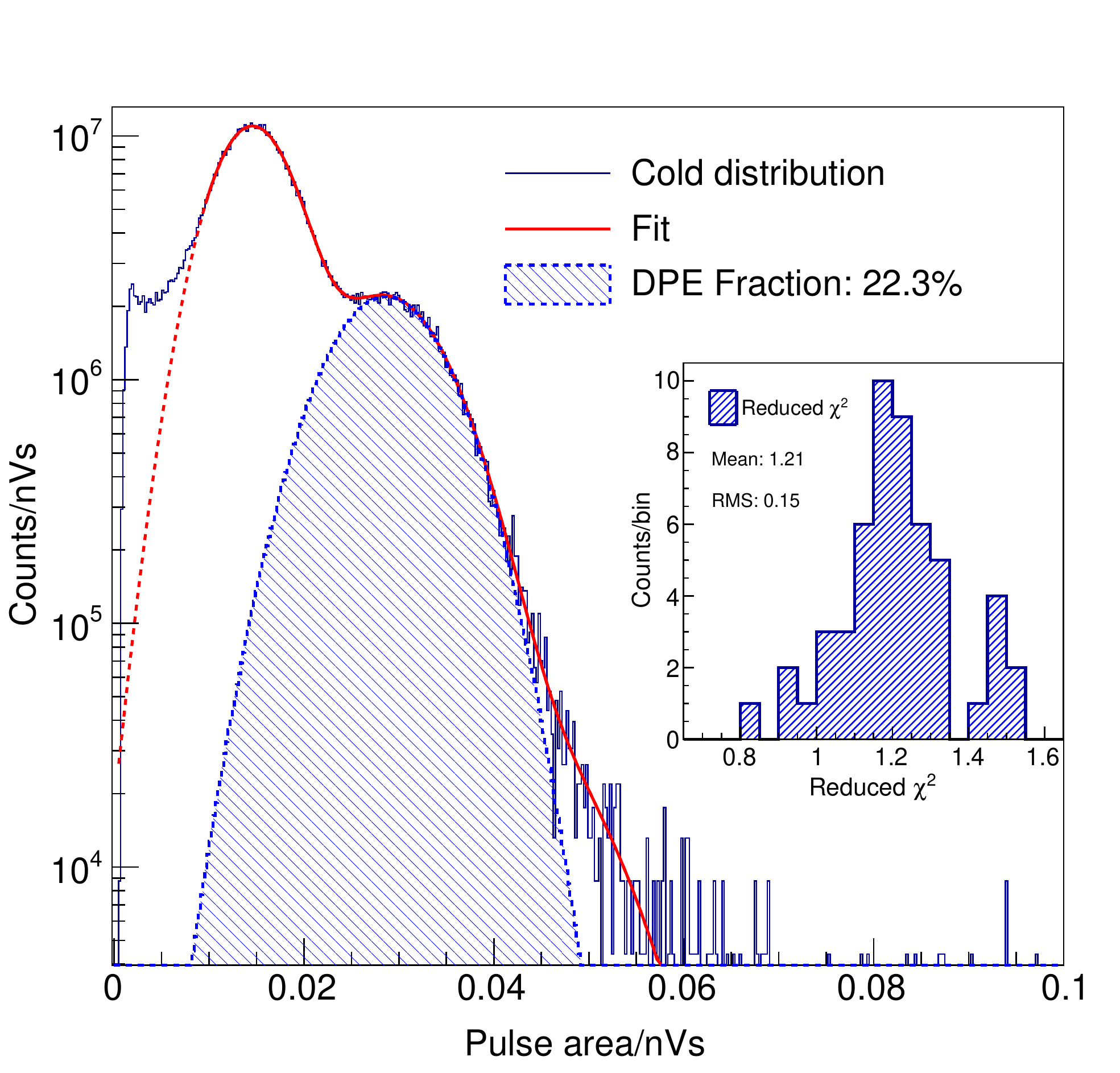}
%\caption{Cold fit}
%\label{fig:ColdFit}
\end{subfigure}
\caption{Double-Gaussian fit of the single VUV photon pulse area distribution at room temperature (left) and at $-97.4^\circ\textrm{C}$ (right). The distributions of reduced $\chi^2$ for all the respective fits are shown as insets.}
\label{fig:VUVgainFitExample}
\end{figure*}

Typical fits of the warm and cold single detected photon pulse area distributions are shown in Figure~\ref{fig:VUVgainFitExample} using the Gaussian model within Equation~\ref{eq:fitmodel}. It was expected that $\sigma^{\scriptscriptstyle\textrm{DPE}} = \sqrt{2}\sigma^{\scriptscriptstyle\textrm{SPE}}$, but the fits yielded a sample standard deviation of $\sigma^{\scriptscriptstyle\textrm{DPE}} = (0.90\pm0.06)\sqrt{2}\sigma^{\scriptscriptstyle\textrm{SPE}}$. This is likely due to a broadening of the SPE distribution due to secondary effects such as: elastic or inelastic back-scattering of electrons off the first dynode and other sources of imperfect multiplication; signals from VUV photon direct hits to the first dynode (see e.g. Ref.~\cite{latepulses, latepulses2} for studies of such effects); and photoelectrons skipping the first dynode~\cite[p.~170]{PMTBOOK2}. A small correlation between two DPE electrons may also contribute. Nonetheless, the fit quality is consistently high for the whole sample, as confirmed by the reduced $\chi^2$ distributions shown in inset in the same figure.

\subsection{Double photoelectron emission fraction}

\begin{figure*}[t]
\begin{subfigure}{0.5\textwidth}
\includegraphics[width=1.0\textwidth]{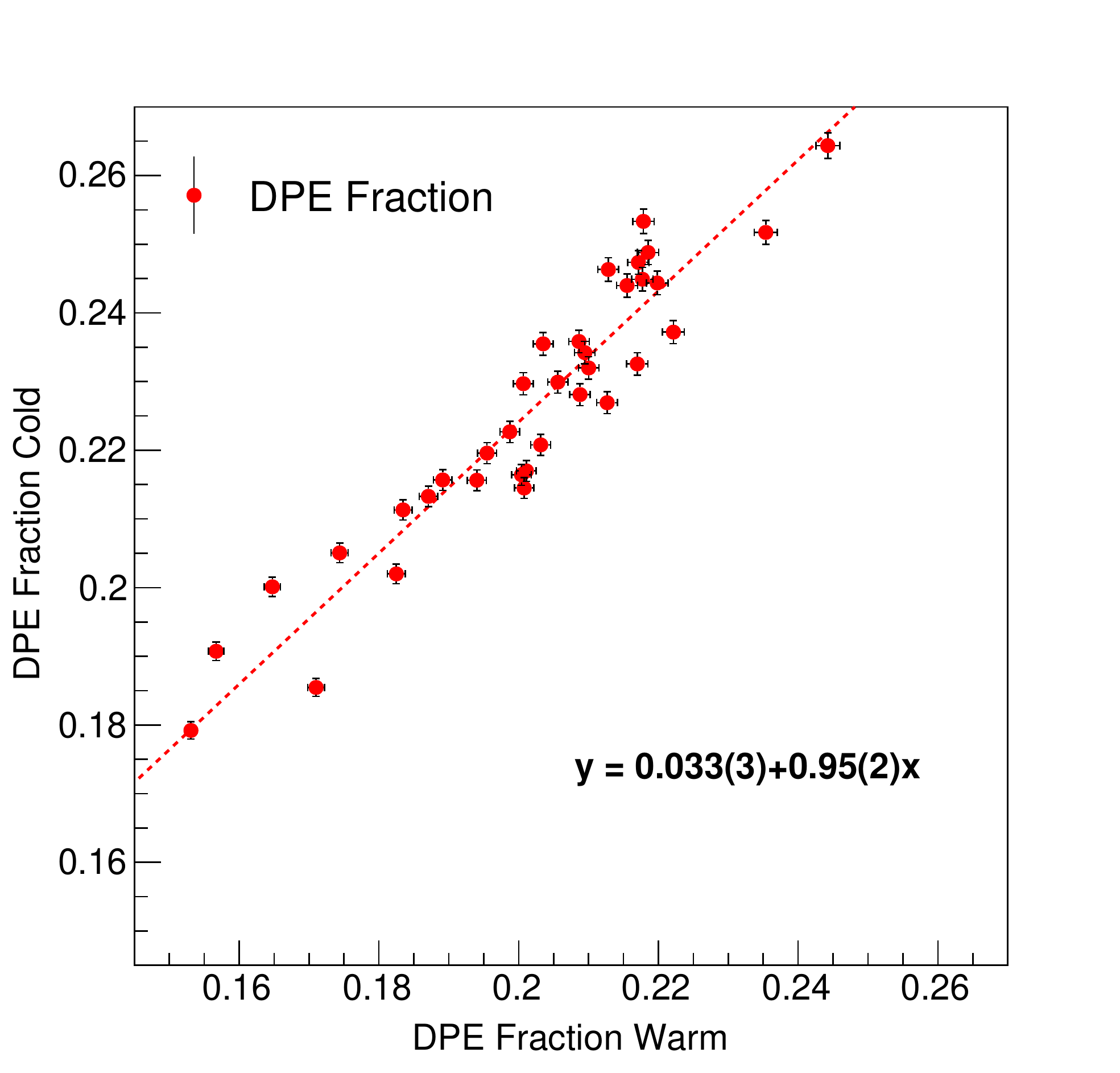}
\end{subfigure}
\begin{subfigure}{0.5\textwidth}
\includegraphics[width=1.0\textwidth]{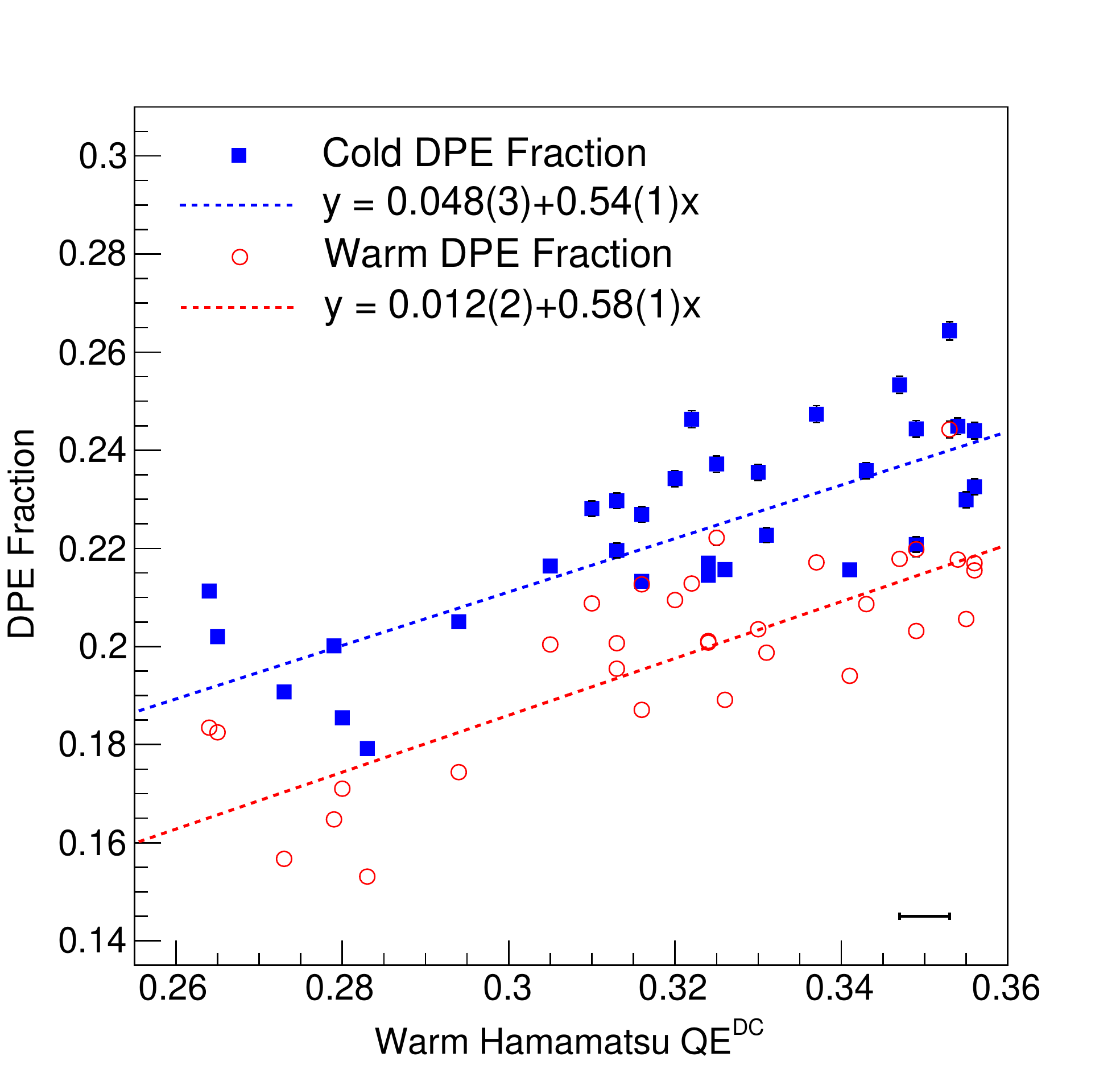}
\end{subfigure}
\caption{Left: DPE fraction measured at $-97.4^\circ\textrm{C}$ plotted against the DPE fraction measured at room temperature for each unit tested. A linear correlation is observed. Right: the DPE fraction measurements are plotted against the $QE^{\scriptscriptstyle\textrm{DC}}_{\scriptscriptstyle\textrm{H}}$ provided by the manufacturer. Linear correlations are also found.}
\label{fig:DPEFractions}
\end{figure*}

The DPE fraction is extracted from the parameter $f^{\scriptscriptstyle\textrm{DPE}}$ in Equation~\ref{eq:fitmodel}. We find that, at room temperature, the 35-PMT population mean and standard deviation are described by $f_{\scriptscriptstyle\textrm{W}}^{\scriptscriptstyle\textrm{DPE}}=(20.2\pm2.0)\%$. Upon cooling, the average fraction increases to $(22.6\pm2.0)\%$---see Figure~\ref{fig:DPEFractions}. Interestingly, a non-zero TPE fraction is also found above the 3-phe background, averaging $\sim\!0.6$\% for all PMTs.

For each individual measurement, the relative uncertainty on $f^{\scriptscriptstyle\textrm{DPE}}$ is $0.7\%$ at the best fit point. Additionally,  there are two leading sources of background that introduce a systematic uncertainty. We already described the contribution of random photon coincidences to the DPE peak in Section~\ref{sec:blueled}. We estimate a $0.1\%$ fraction of VUV photon coincidences. Secondly, dark counts contribute additional fake SPEs. The expected dark rate~\cite{Barrow2017} for these PMTs is shown in Table~\ref{tab:DarkRate}. The two phenomena have the opposite effect on the calculation of the DPE fraction; combined, they represent $<\!0.3\%$ of $f^{\scriptscriptstyle\textrm{DPE}}$, which we consider as a systematic uncertainty since we do not measure the dark rate individually. Muons interacting in the PMT windows and cryostat viewports generate multiple-photon events via Cherenkov emission, but less than 1 muon event per PMT is expected during each measurement, so the effect is negligible. The total relative uncertainty on $f^{\scriptscriptstyle\textrm{DPE}}$ is thus $0.8\%$.

It should be noted that alternative SPE response models can yield different DPE fractions. Instead of the double-Gaussian model assumed above, there may be a physical motivation to choose the Polya distribution, generally used to describe electron multiplication processes. However, we have found that a double-Polya model fits the single photon pulse area distribution of these PMTs only if convolved with a Gaussian of significant width (probably due to the broadening effects discussed in the previous section). The DPE fractions thus obtained are $\sim3$~percentage points lower than the fraction reported above.
The double-Gaussian model was found to be both an accurate and simple representation of the R11410 single VUV photon distribution, easily replicable in other analyses. Hence, it was our model of choice.

\begin{table}[t]
\centering
\caption{Mean number of dark counts expected in a $10$-$\mu$s period for different PMT dark rates, used in the calculation of SPE and DPE population ratios. The fraction relative to the SPE population is also given.}
\label{tab:DarkRate}
\begin{small}
\begin{tabular}{ r | c  l  l } \hline
  & dark rate & counts & fraction\\ \hline
  Max warm & 10~kHz & $\sim$0.1 & $\sim0.65\%$   \\ 
  Typical & 4~kHz & $\sim$0.04 & $\sim0.25\%$  \\ 
  Max cold & 200~Hz & $\sim$0.002 & $\sim0.01\%$ \\ 
  Typical & 40~Hz & $\sim$0.0005 & $\sim0.003\%$ \\ \hline
\end{tabular}
\end{small}
\end{table}

In Figure~\ref{fig:DPEFractions} (left) the DPE fractions at \mbox{$-97.4^\circ$C} are plotted against those measured at room temperature; a clear correlation is found between them. On Figure~\ref{fig:DPEFractions} (right), both warm and cold DPE fractions are plotted against the $QE^{\mbox{\tiny DC}}$ reported by the manufacturer, and a clear correlation is also found in this instance.
The average relative increase of the DPE fraction upon cooling is $12\%$, with a sample standard deviation of $2\%$. This is the first measurement of the temperature dependence of this effect.

\subsection{Quantum efficiency calibration}

\subsubsection{PMT response to main scintillation pulse}

In order to calculate the mean number of photons detected by a PMT due to an $\alpha$-particle interaction in the xenon cell, the total event area is histogrammed and fitted. A Crystal Ball function with mean $b$ and standard deviation $\sigma$ was found to produce good agreement. The mean number of detected photons is then calculated in the asymmetric range $\left[b-3\sigma, b+2\sigma\right]$. Figure~\ref{fig:TypicalAlphaSpectra} shows three typical VUV spectra measured during the testing, with the main $\alpha$-particle interaction plotted as a shaded area for the reference PMT. The area is then calibrated using the $\mu$ value as defined in Section~\ref{sec:theory}.

Determining the precise number of photons reaching the PMT photocathodes requires knowledge of their first-dynode collection efficiency, $\eta$. We have not made a systematic measurement of this parameter, but there are several simulations and measurements in the literature~\cite{Lung2012, Barrow2017, HamamatsuPrivate}. In our tests the PMTs were operated at common gain, using the voltages specified by the manufacturer. However, the behaviour of $\eta$ with respect to $V_\textrm{k-dy1}$ has not been published for this PMT model, so we assumed that $\eta$ was approximately equal for all the PMTs tested. For a distribution of $\eta\simeq0.85\pm0.05$, this introduces a $3\%$ systematic error in the QE measurement later on (we return to this issue in Section~\ref{sec:discussion}).

\begin{figure}[t]
    \centering
    \includegraphics[width=0.5\textwidth]{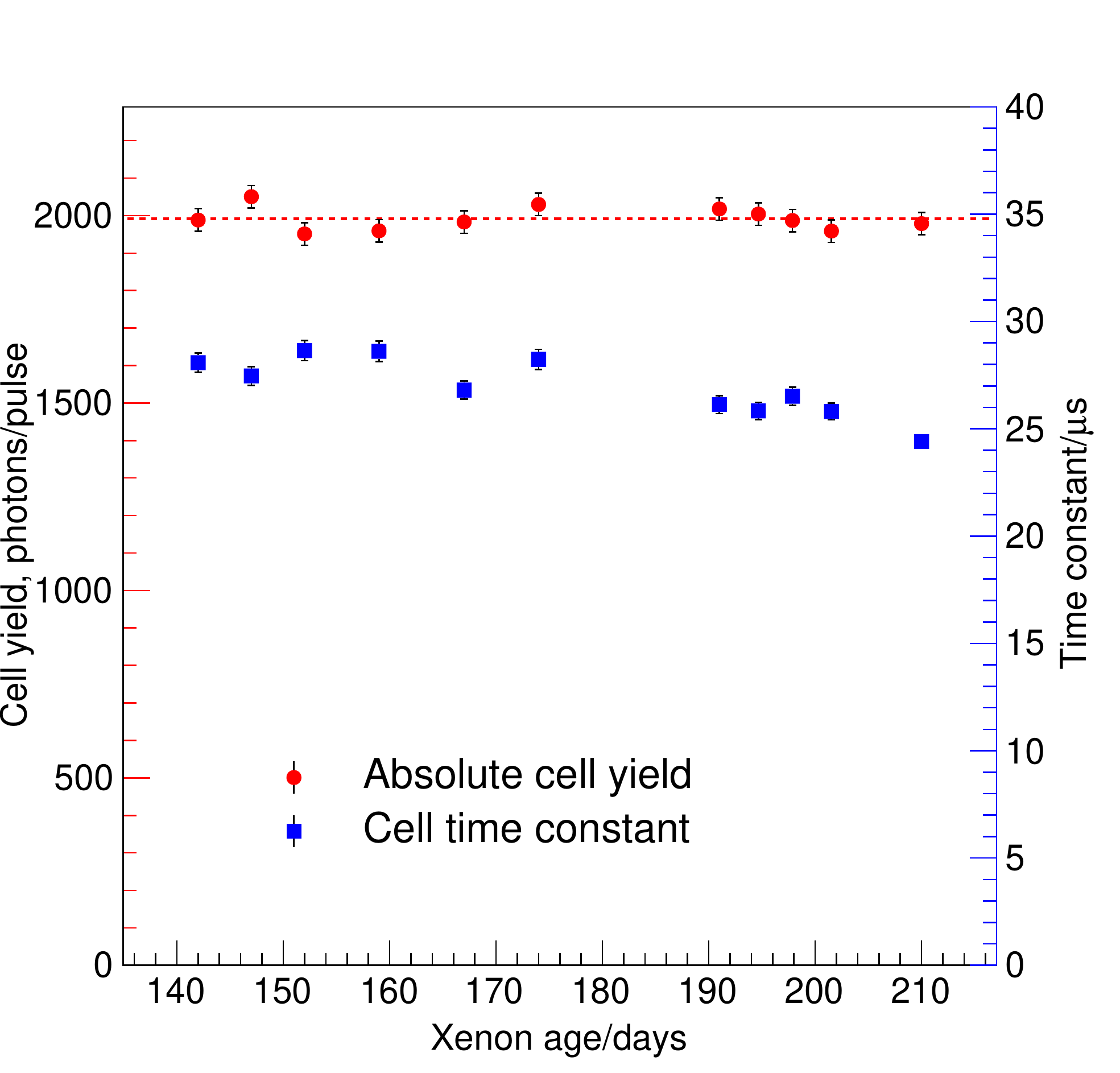}
    \caption{Temporal evolution of the VUV yield of the scintillation cell as measured by the reference PMT assuming $\eta=90\%$ (red circles). Evolution of the mean recombination time for the cell is also shown (blue squares).}
    \label{fig:CellYield}
\end{figure}

The mean number of photons incident on the central (reference) PMT was calibrated as in Equation~\ref{eq:nabs}, with $QE_{\scriptscriptstyle\textrm{H}}^{\scriptscriptstyle\textrm{DC}}$ at $175$~nm ($4$~nm FWHM). The evolution of the cell yield is shown in Figure~\ref{fig:CellYield}, assuming $\eta=90\%$ (see Equation~\ref{eq:etaVUV}). An average of $1992\pm9$~photons per pulse was found for the period of this study. This result agrees with the measurement of a second reference PMT placed in the central position. A variability of $1.5\%$ was observed for the reference measurements. Several tests were carried out to determine its origin. The two reference PMTs were both tested under different magnetic field conditions by rotating the cryostat by $90^\circ$ between consecutive datasets. The PMT gains were observed to change by $+1.5\%$ and $-1.1\%$, respectively, and their QE changed by $+1.4\%$ and $+0.4\%$. This is consistent with the results reported in \mbox{Ref.~\cite{Lung2012}} for a magnetic field of the same magnitude, confirming that the Earth's field does affect the PMT response by a small but measurable amount. We were unable to control for this parameter in our study, but it is unlikely that this would explain all of the variability observed. Another test was performed in which the cell temperature was increased by a few degrees. We acquired datasets at $23.5^\circ\textrm{C}$ and at $30.4^\circ\textrm{C}$ and found a negligible difference of $<\!1\textrm{~phd/$^\circ$C}$. Being unable to correct this residual variability, the full $1.5\%$ was taken as the systematic error on the mean number of VUV photons per pulse.

Likewise, the average flux arriving at the off-axis positions was calibrated using all the off-axis PMTs and then cross-checked with the two reference PMTs. A $1\%$ correction for the thermal contraction of the PMT holder was estimated and applied to the cold measurements. Additionally, a $\lesssim\!2\%$ asymmetry on the light distribution over the array was found. Multiple effects can contribute to this which add up to a few percent systematic error, such as: cell collimator or reflector misalignment, IV misalignment or tilting, MgF$_2$ birefringence, PMT support tilting, and magnetic field effects. It was not possible to correct for these effects to a useful level of precision, so an additional $1\%$ systematic uncertainty was applied to the number of detected photons on the off-axis PMTs.

\begin{figure*}[t]
\begin{subfigure}{0.5\textwidth}
\includegraphics[width=1.0\textwidth]{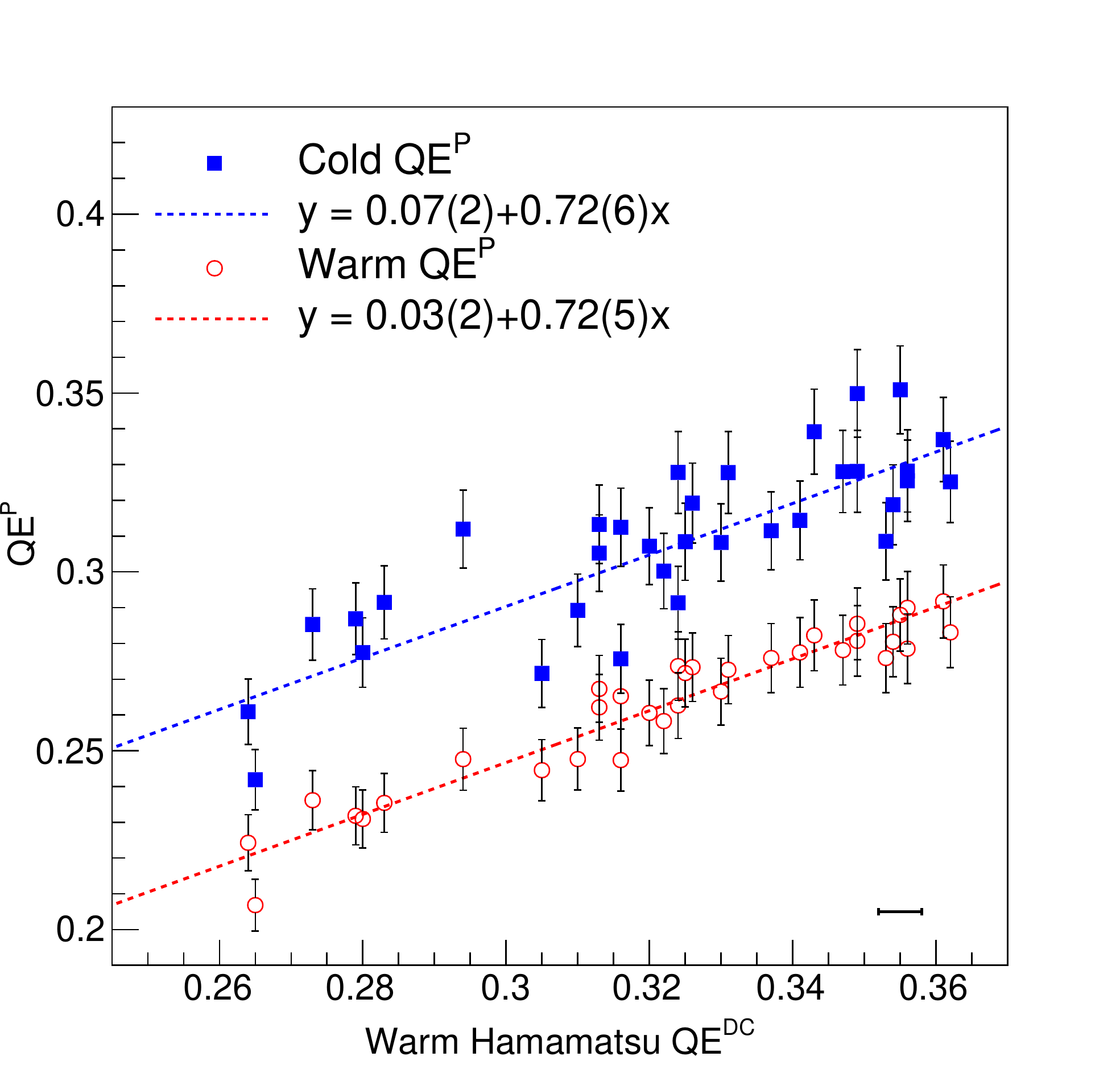}
\end{subfigure}
\begin{subfigure}{0.5\textwidth}
\includegraphics[width=1.0\textwidth]{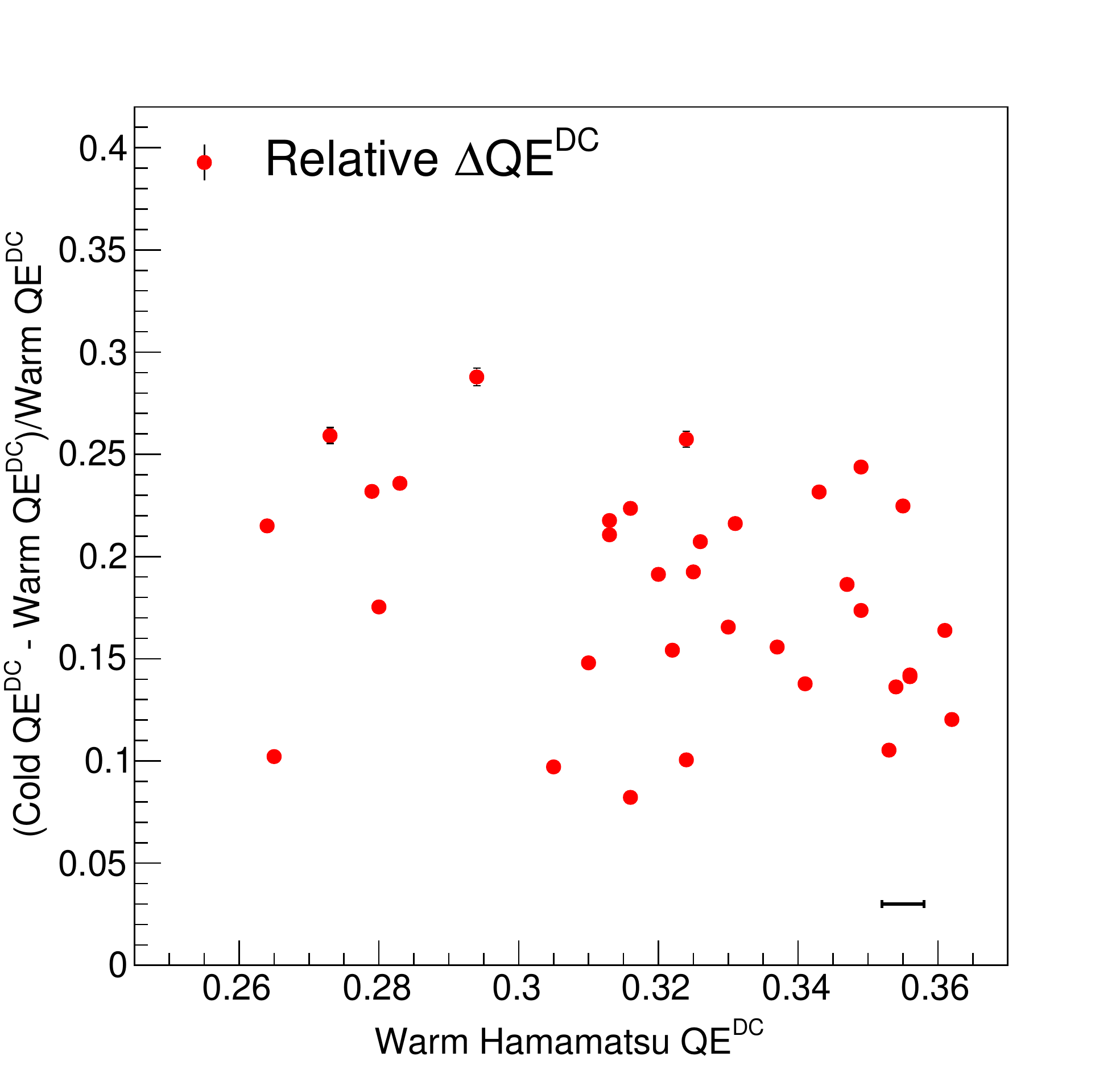}
\end{subfigure}
\caption{Left: $QE^{\scriptscriptstyle\textrm{P}}$ as a function of $QE^{\scriptscriptstyle\textrm{DC}}$ for the warm (red circles) and cold measurements (blue squares). The horizontal line on the bottom right is the systematic uncertainty on $QE^{\scriptscriptstyle\textrm{DC}}$ quoted by the manufacturer. Right: Relative change in $QE^{\scriptscriptstyle\textrm{DC}}$ upon cooling as a function of $QE^{\scriptscriptstyle\textrm{DC}}_{\scriptscriptstyle\textrm{H}}$ for this set of LZ PMTs. No clear correlation is observed.}
\label{fig:QEndeltaQE}
\end{figure*}

The absolute \textit{photocathode} QE is then estimated as $QE^{\scriptscriptstyle\textrm{P}} \!=\! n_{\scriptscriptstyle\textrm{phd}}/\eta\,N$ (see Equations~\ref{eq:nabs} and~\ref{eq:nphd}), assuming $\eta$ is similar for all PMTs. Combining the systematic uncertainties arising from the VUV flux measurements, the small flux asymmetry and the $\eta$ ratio yields a total \textit{relative} systematic error of $3.5\%$ for the QE measurement. $QE^{\scriptscriptstyle\textrm{P}}$ is plotted on Figure~\ref{fig:QEndeltaQE}~(left) as a function of $QE^{\scriptscriptstyle\textrm{DC}}$ for the warm and cold tests. A linear behaviour is found, with the equation indicated in the figure.
Similarly, combining Equations~\ref{eq:nabs} and~\ref{eq:nphe}, an estimate of $QE^{\scriptscriptstyle\textrm{DC}}$ can be made at low temperature. This calibration double-counts DPE pulses but, unlike the manufacturer's DC measurement, it is an average over the whole photocathode. We find $QE_{\scriptscriptstyle\textrm{C}}^{\scriptscriptstyle\textrm{DC}} = (0.04\pm 0.01)+(1.04\pm 0.03)QE^{\scriptscriptstyle\textrm{DC}}_{\scriptscriptstyle\textrm{H}}$.

The relative QE improvement upon cooling has a mean and standard deviation of $(16.9\pm3.9)\%$. For $QE^{\scriptscriptstyle\textrm{DC}}$ the result is $(17.9\pm5.2)\%$. Upon calculating the relative increase, the dependencies on $\eta$ and the flux symmetry cancel out for each PMT, reducing the relative error on each individual measurement to $1.5\%$. The relative change in $QE_{\scriptscriptstyle\textrm{H}}^{\scriptscriptstyle\textrm{DC}}$ is plotted in Figure~\ref{fig:QEndeltaQE} (right) as a function of $QE^{\scriptscriptstyle\textrm{DC}}_{\scriptscriptstyle\textrm{H}}$. No clear correlation is observed.

\subsection{DPE fraction in the ETEL D730/9829QB PMT}
\begin{figure*}[tbh]
 \begin{subfigure}{.5\textwidth}
 \centering 
 \includegraphics[width=\textwidth]{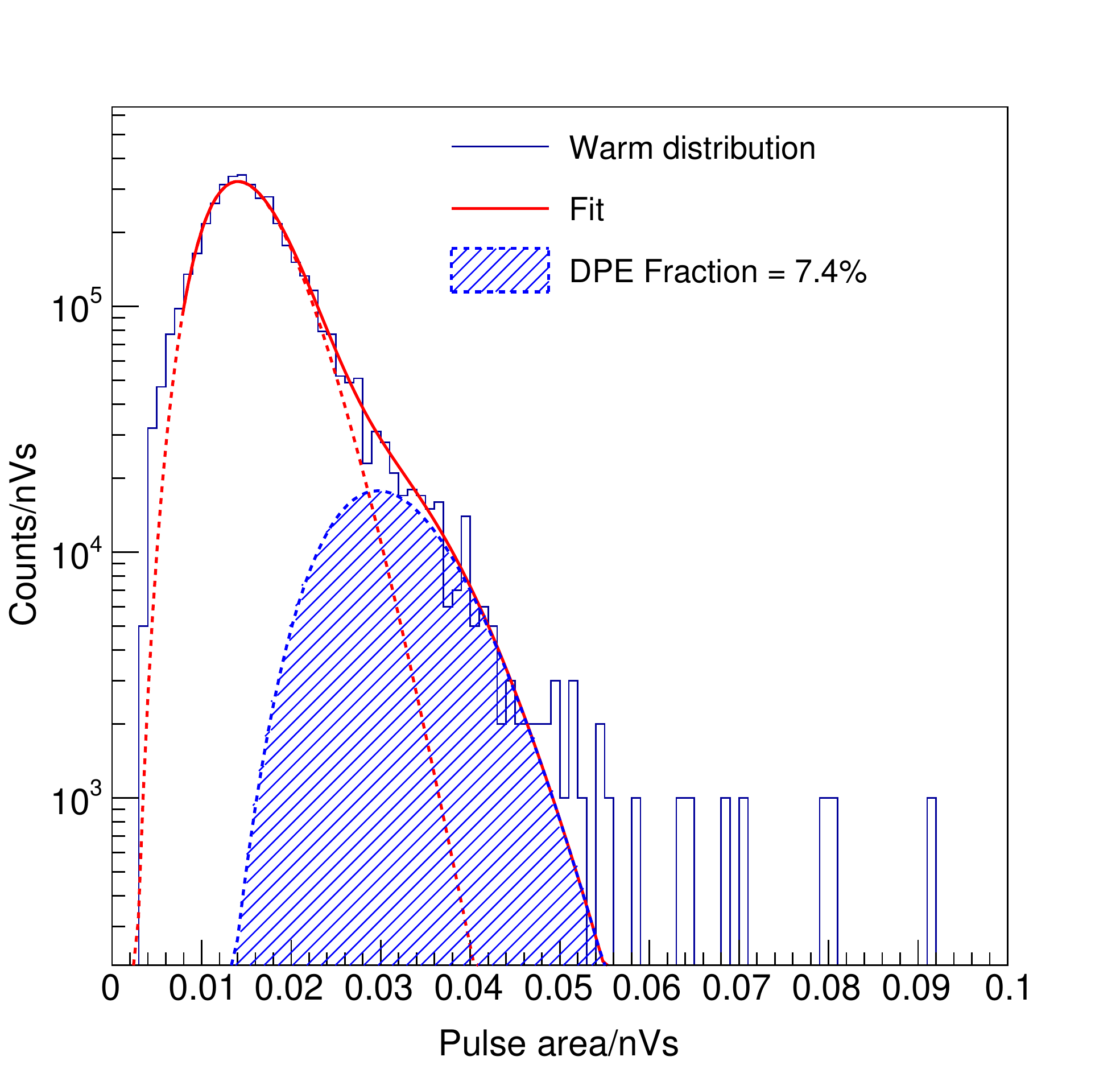}
 \end{subfigure}
 \begin{subfigure}{.5\textwidth}
 \centering
 \includegraphics[width=1\textwidth]{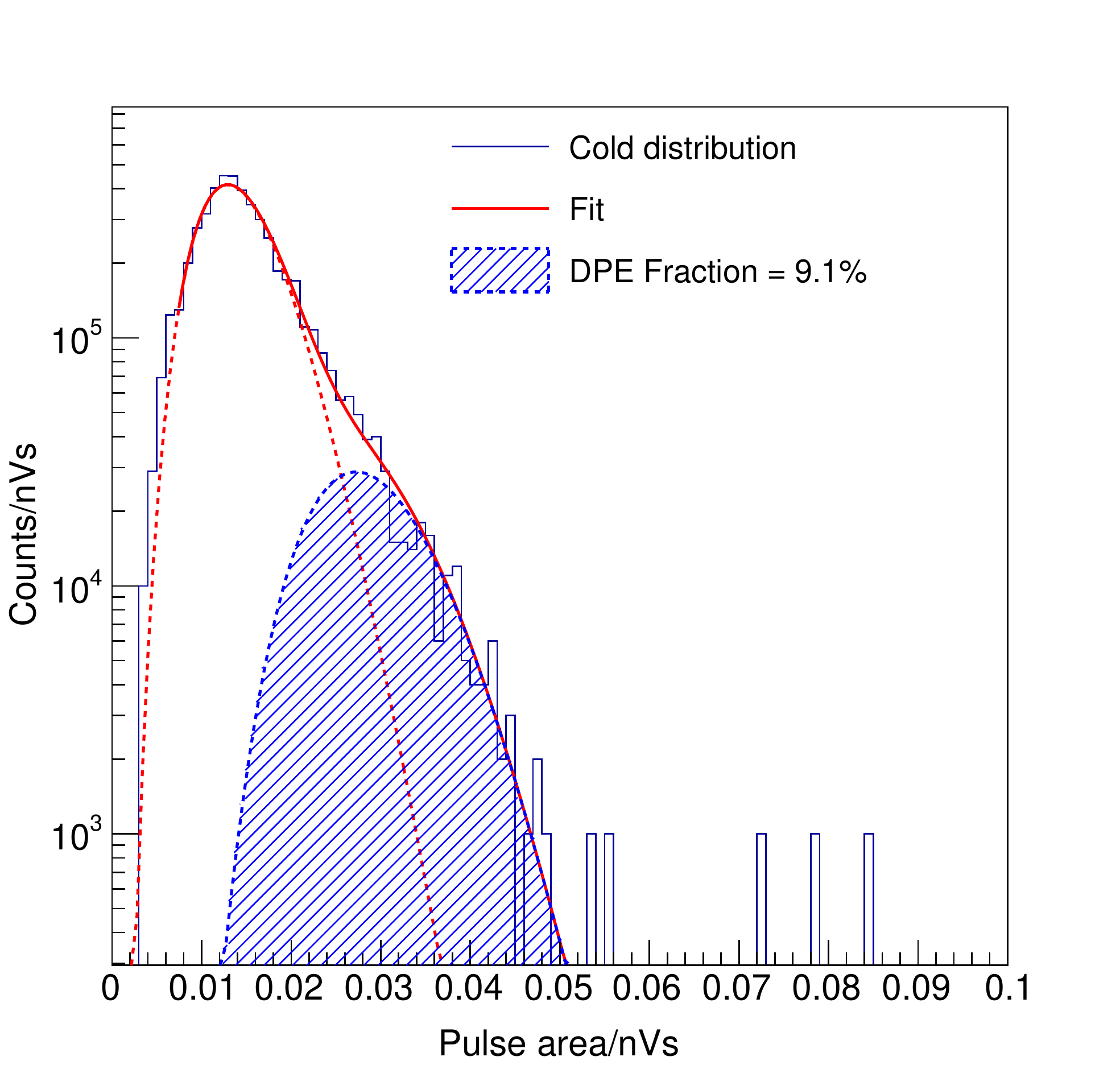}
 \end{subfigure}
 \caption{Double-Polya fit of the single photon pulse area distribution of an ETEL D730/9829QB PMT at room temperature (left) and at $-97.4^\circ\textrm{C}$ (right). A small DPE fraction increase is observed upon cooling, with the cold fraction being $9.1\%$ for the only unit tested.}
 \label{fig:Z3DPEf}
\end{figure*}

The ETEL D730/9829QB was placed in the central cryostat position with the two reference R11410-22 PMTs also in the array. The same trigger settings and similar analysis procedures were adopted, but decreasing the maximum value of $\tau$ to $6.5$~ns and enforcing an inhibit period of 3~$\mu$s prior to each pulse to prevent selection of afterpulses, which are more prevalent in this $>$10-year old PMT.

As described in Section~\ref{sec:theory}, we used the double Polya model to fit the single photon pulse area of this PMT model. The results are shown on Figure~\ref{fig:Z3DPEf}. The fit is consistent with a value of $f_{\scriptscriptstyle\textrm{C}}^{\scriptscriptstyle\textrm{DPE}}\simeq\!10\%$ at $-97.4^\circ\textrm{C}$ and a small increase is observed upon cooling.

%%%%%%%%%%%%%%%%%%%%%%%%%%%%%%%%%%%%%%%%%%%%%%%%%%%%%%%%%%%%%%%%%%%%%%%%%%%
\section{Discussion}
\label{sec:discussion}

In the ensuing discussion we treat the photocathode response in the context of the simple three-step model for semiconductor crystals mentioned in Section~\ref{sec:theory}~\cite{3stepmodel}.

\subsection{Hamamatsu R11410-22}

In our sample of 35 R11410-22 PMTs we find that $QE^{\scriptscriptstyle\textrm{DC}}_{\scriptscriptstyle\textrm{H}}$ provides a consistent measurement across the whole population, despite probing only a $5$~mm spot size at the centre of the PMT window. We have confirmed that the PMT QE improves upon cooling, with results generally consistent with previous measurements for this model~\cite{Lyashenko2014} and also consistent with the ZEPLIN-III PMT model~\cite{Araujo2004}. We have measured the DPE fraction under VUV illumination at room temperature, obtaining values in agreement with Ref.~\cite{Faham2015}. We confirm that the effect is negligible under \textit{blue} illumination ($435$~nm). We have also measured for the first time the increase in DPE fraction between room temperature and $\simeq -100^\circ\textrm{C}$.

A very small TPE fraction of $\sim\!0.6\%$ is observed in our data, and this is consistent with the impact ionisation explanation that also accounts for the DPE phenomenon. This behaviour has long been observed in bialkali photocathodes~\cite{Sobieski1976,Johnson1988} and it lends a particular shape to the PMT spectral response. Under visible light illumination only single photoelectron signals are observed. For decreasing wavelengths the impact ionisation probability rises, but the QE will slowly decrease as neither electron in the pair may have enough energy to reach the vacuum. For sufficiently short wavelengths, a response can be measured even if impact ionisation has occurred (as now both electrons have higher energy) and the QE rises again; in some cases both electrons are emitted creating a DPE response. Further decreasing the wavelength leads to a similar oscillatory behaviour of the QE, a gentle decrease followed by an increase, and eventually to a sizeable TPE fraction, and so on. The initial reduction in QE observed in Refs.~\citep{Sobieski1976,Johnson1988} for wavelengths just shorter than xenon scintillation is probably caused by an increase in impact ionisation, rather than by the spectral cut-off of the quartz window; this conclusion is supported by the presence of a small TPE fraction in our data and by the spectral behaviour of $QE^{\scriptscriptstyle\textrm{DC}}_{\scriptscriptstyle\textrm{H}}$ reported by the manufacturer---which decreases faster than expected for the synthetic silica transmission spectrum. A similar behaviour was observed in Ref.~\cite{Sobieski1976}.

With respect to the PMT performance with temperature there is, as expected, a clear linear relationship between the room temperature QE and that at $-97.4^\circ\textrm{C}$. In turn, this translates into a very weak correlation between the relative increase in $QE^{\scriptscriptstyle\textrm{DC}}$ and its value at room temperature. The trend that lower QE PMTs improve the most upon cooling reported for the ZEPLIN-III PMTs in Ref.~\cite{Araujo2004} is, on first inspection, much more pronounced than that measured here. However, we note that the more marked behaviour was seen for the lower QE PMTs, and our Hamamatsu sample has in general only high QE units. The two trends are in fact compatible if one considers only those PMTs with $QE^{\scriptscriptstyle\textrm{DC}}>25\%$ in Ref.~\cite{Araujo2004}.

As shown in Figure~\ref{fig:DPEFractions}, the DPE fraction in the R11410-22 model is correlated with $QE^{\scriptscriptstyle\textrm{DC}}_{\scriptscriptstyle\textrm{H}}$, which is a straightforward consequence of the DPE effect. The QE measurement performed by Hamamatsu involves the ratio between the DC photocurrent measured directly from the photocathode and the incident VUV flux from a filtered deuterium lamp~\cite{HamamatsuPrivate} calibrated by a reference sensor. This ratio is directly affected by the fraction of DPE events, and $QE^{\scriptscriptstyle\textrm{DC}}_{\scriptscriptstyle\textrm{H}}$ is therefore expected to be equal to $(1+f^{\scriptscriptstyle\textrm{DPE}}_{\scriptscriptstyle\textrm{W}})\cdot QE^{\scriptscriptstyle\textrm{P}}_{\scriptscriptstyle\textrm{W}}$.

The correlation is weaker, but still present, with respect to $QE^{\scriptscriptstyle\textrm{P}}$. In this instance we may infer the following explanation. A higher QE is both due to the probability of a photon interacting in the photocathode (e.g.~due to its precise thickness) and the probabilities for the photoelectron to diffuse to the surface and to be emitted from the material. Drifting electrons scatter off phonons and lattice imperfections, and thus the photocathode QE is influenced by the quality of the sensitive layer and the temperature. Thus, the two electrons in a DPE event are both more likely to be emitted given a higher $QE^{\scriptscriptstyle\textrm{P}}$ measurement.

Nonetheless, we also observe that the average relative increase in DPE fraction with cooling is lower than that experienced by $QE^{\scriptscriptstyle\textrm{P}}$. We suggest that this could be a consequence of the lower average energy of DPE electrons. The reduction of the electron-phonon scattering rate upon cooling increases the mean free path of all electrons, but the lower energy of DPE electrons means that often at least one of them will not diffuse to the surface of the photocathode or overcome the work function to make it into the vacuum. We conclude that the increase in impact ionisation in the VUV region can enhance $QE^{\scriptscriptstyle\textrm{P}}$ by making available two electrons with higher combined probability of escaping from the material relative to a single photoelectron.

We now turn to the issue of first dynode collection efficiency. Several measurements and simulations of this parameter have been made for this PMT model~\cite{Lung2012, Barrow2017, HamamatsuPrivate}. At our operating voltages, $\eta \simeq 85\%$ is the value indicated by the manufacturer. To our knowledge, only visible light sources have been used to characterise $\eta$ for these PMTs, while a wavelength-dependent behaviour has been observed at least in some models---in particular, this parameter has been known to increase for UV wavelengths and the reason has not been understood previously~\cite{PMTBOOK}. We propose that this behaviour can be attributed to double photoelectron emission. In a DPE event, two electrons are available to overcome this inefficiency, and hence the collection efficiency specifically for DPE events, $\eta_{\scriptscriptstyle\textrm{DPE}}$, is expected to be very high, certainly higher than that measured when only one photoelectron is involved.

For a simple estimate of this effect we may assume that each DPE electron has a probability $\eta_{\scriptscriptstyle\textrm{SPE}}$ of being multiplied, increasing the average efficiency to 88\% for VUV illumination. It should be noted that the use of the measured DPE fraction neglects some genuine DPE electrons emitted from the photocathode that do not overcome the collection efficiency. In this instance, an improved estimate for this parameter is 
\begin{equation}
\label{eq:etaVUV}
\eta_{\scriptscriptstyle\textrm{VUV}} = \eta_{\scriptscriptstyle\textrm{SPE}} + (1-\eta_{\scriptscriptstyle\textrm{SPE}})(f^{\scriptscriptstyle\textrm{DPE}}/\eta^2_{\scriptscriptstyle\textrm{SPE}})\eta_{\scriptscriptstyle\textrm{SPE}}\textrm{.}
\end{equation}
Assuming $\eta_{\scriptscriptstyle\textrm{SPE}}\simeq0.85$ and $f^{\scriptscriptstyle\textrm{DPE}}\simeq0.23$, we estimate $\eta_{\scriptscriptstyle\textrm{VUV}}\simeq0.89$.

In both calculations we have made two assumptions: that the collection efficiency is independent of electron energy at emission from the photocathode, and that the efficiencies for the two DPE electrons are uncorrelated. While the latter assumption is at least partially true for some relevant parameters, namely the initial transverse momentum or the elastic (or partially elastic) reflections off the first dynode, others such as the emission location on the photocathode are certainly correlated. In any case, it is reasonable to expect a slight increase in collection efficiency at VUV wavelengths, as suggested by some studies, given a sizeable DPE fraction. A precise measurement of $\eta$ for the R11410 model is desirable to understand precisely the response to xenon scintillation light and other short-wavelength measurements.

\subsection{ETEL D730/9829QB}

The measurement of a non-zero DPE fraction in the ETEL D730/9829QB model used in ZEPLIN-III deserves some comment. This is, to our knowledge, the only model studied for this effect which was not manufactured by Hamamatsu. Two distributions are presented in Figure~2 of Ref.~\cite{Santos2012} for the (VUV) electroluminescence response caused by single electrons emitted from the liquid, obtained in ZEPLIN-III using these PMTs---including the actual unit tested in this work. The mean of the distribution reported using photon counting is 11\% smaller than that obtained by pulse area integration normalised to the single photoelectron response calibrated from thermionic dark counts in the same dataset. This discrepancy is generally consistent with the DPE effect of $(9.1\pm0.1)\%$ measured for the ETEL PMT. The effect is, in any case, noticeably smaller than that in the R11410 PMTs. We note that Hamamatsu employs a more recent bialkali photocathode technology optimised for low temperature operation \cite{Nakamura2010}, which is likely to differ in several respects (composition, thickness, and crystalline properties) from the bialkali employed by ETEL.

Clearly, different bialkali photocathode technologies impact the measurable DPE fraction. Another example is provided by the Hamamatsu R8778 PMTs used in the LUX experiment. LUX reports an $\sim\!18\%$ DPE fraction for an average $QE^{\scriptscriptstyle\textrm{DC}}_{\scriptscriptstyle\textrm{H}}\simeq33\%$~\cite{Faham2015, LUXInst}. For these QE values, similar to those of our sample, the R11410 photocathode exhibits a higher DPE (Figure~\ref{fig:DPEFractions}, right).

In ZEPLIN-III the calibration of the single photoelectron response relied on the method described in Ref.~\cite{Neves2010}, which is unaffected by the DPE effect. This starts with the calculation of the mean number of detected photons from the frequency of zero response for small signals, leading to a linear plot for $\mu_1$ versus mean pulse area. In the presence of double emission with constant probability $f_{\scriptscriptstyle\textrm{DPE}}$, the mean area obtained in response to single VUV photons is thus $\mu_1(1+f_{\scriptscriptstyle\textrm{DPE}})$, which confirms that this effect was correctly accounted for in the various ZEPLIN-III analyses in terms of mean response---if not in signal variance.

%%%%%%%%%%%%%%%%%%%%%%%%%%%%%%%%%%%%%%%%%%%%%%%%%%%%%%%%%%%%%%%%%%%%%%%%%%%
\section{Conclusions}
\label{sec:conclusions}

We have calibrated 35 R11410-22 photomultiplier tubes for the LZ experiment under VUV illumination provided by a xenon scintillation cell at room temperature and at $-97.4^{\circ}$~C. We have measured the double photoelectron emission fraction at both room and low temperature and found them to correlate linearly with the $QE^{\scriptscriptstyle\textrm{DC}}$ parameter measured by the manufacturer. The DPE effect has been studied as a function of temperature for the first time. The average DPE fractions were found to be $(20.2\pm2.0)\%$ and $(22.6\pm2.0)\%$ warm and cold, respectively (uncertainties correspond to the sample standard deviation). The average $QE^{\scriptscriptstyle\textrm{DC}}$ of this sample was $32.4\%$. We estimated an average improvement $\Delta QE^{\scriptscriptstyle\textrm{DC}} = (17.9\pm5.2)\%$ upon cooling, with a weak correlation with the room temperature value. We have estimated the response to the VUV xenon scintillation light and found $QE^{\scriptscriptstyle\textrm{P}}$ to be on average $(26.5\pm2.2)\%$ (warm) and $(30.9\pm2.5)\%$ (cold)---the latter is the relevant parameter needed for liquid xenon photon counting experiments such as LZ.

Thus, the presence of double photoelectron emission at VUV wavelengths---and the suggested increase in first dynode collection efficiency---should be taken into account in the analysis of data from liquid and gaseous xenon detectors. It is also worth pointing out that these factors may have affected, to a small extent, various historical measurements of the optical properties of materials and devices at very short wavelengths.

\section*{Acknowledgements}
The authors would like to thank Hamamatsu Photonics for useful discussions and information on the R11410 photomultiplier. We wish to thank our collaborators in the LZ Dark Matter Experiment for their ongoing support---in particular to our colleagues at Brown University, where the PMT acceptance testing programme for LZ is taking place, and to T.~Biesiadzinski, C.~Hall, F.~Neves and P.~Sorensen for providing comments on this manuscript. Thanks are due to A.~Rochester from the Physics Instrumentation Workshop at Imperial and to R.~Preece at RAL for helping with the mechanical design; to the Physics and the HEP group workshops for assisting with fabrication; to the Physics and HEP group electronics workshops for their assistance; and to Kurt~J.~Lesker and Allectra for delivering the cryostat hardware. We thank F.~Neves and V.~Solovov at LIP-Coimbra for advice on data acquisition and analysis, and Imperial undergraduates J.~Lo, A.~Sahota and S.~Rai for their assistance with thermal and optical modelling. F.~Froborg has received funding from the European Union's Horizon 2020 research and innovation programme under the Marie Sklodowska-Curie grant agreement No 703650. This work was supported by the U.K. Science and Technology Facilities Council (STFC) under award number ST/M003655/1.
%\newpage

\FloatBarrier
\section*{References}

\bibliography{elsarticle-template}

\end{document}